\newif\ifShowKeys
\ShowKeysfalse

\documentclass[11pt]{article}
	\pdfoutput=1
\usepackage[usenames,dvipsnames]{xcolor}
\usepackage[setpagesize=false,pagebackref=false, linktocpage, bookmarksopen=true, colorlinks=true, linkcolor=Maroon,citecolor=Maroon,urlcolor=Maroon]{hyperref}
\usepackage[parsep]{collref}

\ifShowKeys \usepackage[notcite]{showkeys} \fi

\usepackage{amsmath, amssymb,amsthm}
\numberwithin{equation}{section}

\usepackage{bm,environ,mathrsfs,array,arydshln}
\usepackage{booktabs,float,slashed,hyperref}
\usepackage[mathcal]{euscript}
\usepackage{tensor} 						% Ratcliffe package to write tensors
\usepackage{mathabx}
\usepackage[vcentermath]{youngtab}
\usepackage{xcolor}

\usepackage{aurical}
\usepackage[T1]{fontenc}
\usepackage[nodayofweek]{date time}
\usepackage{graphicx,epsfig,epic}
\usepackage{tikz} 
\usetikzlibrary{arrows,decorations.pathreplacing,decorations.markings,snakes}
\usetikzlibrary{cd}

\usepackage{framed}						% for shaded equations \begin{shaded}...\end{shaded}
\definecolor{shadecolor}{rgb}{0.9996078, 0.984314, 0.960784}

\usepackage{comment}

\allowdisplaybreaks

\definecolor{myred}{RGB}{233, 33, 45}

\newcommand{\bs}{\begin{shaded}}
\newcommand{\es}{\end{shaded}\noindent}
\def\ba#1\ea{\begin{align}#1\end{align}}		% very clever way to bypass the known problem...
\newcommand{\be}{\begin{equation}}
\newcommand{\ee}{\end{equation}}
\newcommand{\bea}{\begin{equation} \begin{aligned}} 
\newcommand{\eea}{\end{aligned} \end{equation}}
\newcommand{\mc}{\mathcal }
\newcommand{\mb}{\mathbb }
\newcommand{\la}{\label}
\newcommand{\eps}{\varepsilon}

\newcommand{\lp}{\notag \\ & }

\newcommand{\cf}{\textit{cf.} }
\newcommand{\ie}{\textit{i.e.} }
\newcommand{\eg}{\textit{e.g.} }

\newcommand{\N}{\mathcal N}

\def\ov{\over}

\def \b  {\beta}

\def \ha {{1\ov 2}}
 \def \nb {n_{_\b}}

\renewcommand{\l}{\lambda}

\DeclareMathOperator{\Tr}{Tr}

\DeclareMathOperator{\PE}{PE}

\newcommand{\I}{\mathrm{I}}
\newcommand{\Isp}{\widehat{\I}}

\newcommand{\Z}{\mathcal Z}
\newcommand{\Zsp}{\widehat{\Z}}

\newcommand{\T}{{\rm T}}

\newcommand{\sX}{{\sf X}}
\newcommand{\sY}{{\sf Y}}
\newcommand{\sZ}{{\sf Z}}

\newcommand{\ul}[1]{\underline{#1}}
\newcommand{\ulh}[1]{\underline{\hat{#1}}}

\begin{document}

%\date{\currenttime}
%\begin{flushright}\boxed{\small{\tt \today \ \ - \ \  \currenttime }}\end{flushright}

%\centerline{\Large\sc  Schur index and D3 branes - Notes}
%\vskip 0.5cm
%% \centerline{\sc M. Beccaria}
%\bigskip
%\begin{abstract}
%\begin{center}
%%\includegraphics[width=0.5\textwidth]{cover}
%\end{center}
%\end{abstract}

%%%%%%%%%%%%%%%%%%%%%%%%%%%%%%%%%%%%%%%%%%%%%%%%%%%%%%%%%
%%%%% NOTICE the title page is commented out with \begin{comment}...\end{comment}   but is ready to be used
%\begin{comment}

\begin{titlepage}
%\begin{tabbing}
%\hspace*{11.5cm} \=  \kill % set the tabbings
%\>  Imperial-TP-AT-2024-?? \\
%\> %none
%\end{tabbing}

\vspace*{15mm}
\begin{center}
{\Large\sc   Large $N$  Schur index of $\N=4$ SYM}\vskip 9pt
{\Large\sc    from semiclassical  D3 brane}

\vspace*{10mm}

{\Large M. Beccaria${}^{\,a}$, A. Cabo-Bizet$^{\,a, }$}
%\footnote{\ Also at the Institute for Theoretical and Mathematical Physics (ITMP) of Moscow University   and Lebedev Institute.}} 

\vspace*{4mm}
	
${}^a$ Universit\`a del Salento, Dipartimento di Matematica e Fisica \textit{Ennio De Giorgi},\\ 
		and I.N.F.N. - sezione di Lecce, Via Arnesano, I-73100 Lecce, Italy
			\vskip 0.3cm
%${}^b$ Blackett Laboratory, Imperial College London SW7 2AZ, U.K.
%			\vskip 0.3cm
\vskip 0.2cm {\small E-mail: \texttt{matteo.beccaria@le.infn.it, acbizet@gmail.com}}
\vspace*{0.8cm}
\end{center}

\begin{abstract}  
We consider the refined Schur superconformal index of 4d $\N=4$ $U(N)$ SYM and the first term of its giant-graviton expansion, first 
predicted in arXiv:2001.11667 using indirect superconformal algebra considerations and analytic continuation of fugacities. This correction 
is the leading non-perturbative correction to the index at large $N$
and we reproduce it from the semiclassical partition function of quantum D3 brane wrapped on $S^{1}\times S^{3}$ in 
a twisted modification of the $AdS_{5}\times S^{5}$ string background, depending on the index R-symmetry fugacity.
Our calculation does not exploit directly supersymmetry. It is based on the determination of the partition function of the various bosonic and fermionic
fluctuations on the wrapped brane whose action is conformal with specific constant holonomies along thermal cycle. 
We  show how those partition functions may be obtained by adapting the operator counting method of Cardy to the twisted background.  
\end{abstract}
\vskip 0.5cm
	{
		%Keywords: {\sc insert here keywords}
	}
\end{titlepage}
%\end{comment}
%%%%%%%%%%%%%%%%%%%%%%%%%%%%%%%%%%%%%%%%%%%%%%%%%%%%%%%%%

\tableofcontents
\vspace{1cm}

%\section*{To be done}

%\section*{\red{Questions}}
%
%\begin{enumerate}
%\item
%\end{enumerate}

\section{Introduction}

The AdS/CFT correspondence establishes a duality between a boundary conformal gauge theory  and a string dual 
where $N\sim 1/g_{s}$ at constant 't Hooft planar coupling, with $N$ being the gauge group rank and  $g_{s}$ the string coupling.  
A general puzzle  appearing at finite $N$ is that the number of boundary gauge theory states at fixed conserved charges decreases  due to trace relations
while non-perturbative states in the gravity theory become lighter at increasing  $g_{s}$. A way out of this  apparent contradiction is to call on some kind of stringy exclusion principle  
%\cite{Maldacena:1998bw,Myers:1999ps,McGreevy:2000cw}.
\cite{Maldacena:1998bw,McGreevy:2000cw}.

A context where this issue can be addressed is the study of finite $N$ corrections to the superconformal index \cite{Kinney:2005ej} of a supersymmetric gauge theory with gravity dual.
These are corrections that in many instances take the form of a giant graviton-like expansions \cite{Gaiotto:2021xce} and originate from gauge theory states with R-charge of order $N$. 
Let us denote by $\I_{N}(q)$ the superconformal index with $q$ collectively standing for the set of fugacities. Finite $N$ effects admit the (typical) representation \footnote{For instance, the term $q^{kN}$ may be replaced
by $q^{a k N}$ with some constant $a$. Also, the index $k$ is often a multi-index. In the present discussion, these details are not relevant.}
\be
\la{1.1}
\I_{N}(q) = \I_{\rm KK}(q)\, \left[1+\sum_{k=1}^{\infty} q^{kN}\, \I_{k}^{\rm GG}(q)\right]\, ,
\ee
where $\I_{\rm KK}(q) \equiv \I_{\infty}(q)$ 
is the $N\to \infty$ limit of the index and the key property of the quantities $\I_{k}^{\rm GG}(q)$ is that they do not depend on $N$, at least at generic points in 
fugacity space. Thus, (\ref{1.1}) encodes the full finite $N$ dependence of the index as a sum of corrections that are non-perturbative at large $N$.

The naming of the expansion is due to the fact that  finite $N$ corrections to the large $N$ index are captured on gravity side by certain wrapped D-branes BPS large ``giant'' configurations.
This interpretation  was developed on a quantitative level in the works of Imamura and collaborators  
\cite{Arai:2019xmp,Arai:2019wgv,Arai:2019aou,Arai:2020qaj,Arai:2020uwd,Fujiwara:2021xgu,Imamura:2021ytr,Imamura:2021dya,Imamura:2022aua,Fujiwara:2023bdc} (see also \cite{Bobev:2023bxs})
where a systematic analysis of finite $N$ effects in superconformal index from wrapped branes was carried out. In that approach, 
the  factor $\I_{\rm KK}(q)$ is the contribution to the index from supergravity Kaluza-Klein BPS states (hence the label).
The ``classical'' factor $q^{kN}$ comes from the classical action 
of the wrapped brane, with $k$ being a topological wrapping number. The remaining non-trivial pieces $\I_{k}^{\rm GG}(q)$ represent the  index of  fluctuation modes
on the wrapped brane, a non-trivial  superconformal 
theory with BPS reduced supersymmetry. \footnote{
Working on the gauge theory side, apart from AdS/CFT, the expansion (\ref{1.1}) was shown to arise counting invariants in any unitary matrix model \cite{Murthy:2022ien}. The expansion can be 
derived by free-fermion methods and interpreted as an instanton expansion in gauge theory in \cite{Eniceicu:2023cxn}. The first and second terms in this free-fermion expansion were derived in \cite{Murthy:2022ien} and \cite{Liu:2022olj}.
The expansion obtained by this approach is exact but does not agree term by term with the wrapped D-brane expansion. This is not a contradiction because at mathematical level the index itself does not determine $\widehat{\I}_{k}(q)$ in (\ref{1.1}), 
see for instance \cite{Eniceicu:2023uvd}.
}

The relation between (\ref{1.1}) and a possible manifestation of a stringy exclusion principle lies in the fact that  contributions at increasing $k$  have typically alternating signs.
As an illustrative paradigmatic example, the $\frac{1}{2}$-BPS index of 4d $\N=4$ U(N) SYM theory on $S^{1}\times S^{3}$ takes the form (\ref{1.1}) with  ($R$ is a Cartan of the $SU(4)$ R-symmetry and $(q)_{N}$ is the Pochhammer symbol)
%\be
%\I_{N}(q) = \Tr(-1)^{\rm F}q^{R} = \frac{1}{(q)_{N}} = \frac{1}{(q)_{\infty}}\,\sum_{k=0}^{\infty}q^{kN}\, (-1)^{k}\frac{q^{k(k+1)/2}}{(q)_{k}}\, ,
%\ee
\be
\la{1.2}
\I_{N}^{\rm SYM}(q) = \Tr(-1)^{\rm F}q^{R} = \frac{1}{(q)_{N}}, \qquad \widehat{\I}_{k}(q)  = (-1)^{k}\frac{q^{k(k+1)/2}}{(q)_{k}}\, ,
\ee
where the trace is restricted to $\frac{1}{2}$-BPS states. 
The factor $(-1)^{k}$ inside $\widehat{\I}_{k}(q)$ 
suggests the emergence of an extra bulk grading  \cite{Lee:2022vig,Lee:2023iil}. As remarked in \cite{Eleftheriou:2023jxr},
this grading is closely related to the fact that the giant graviton expansion from the bulk is an Euclidean functional integral calculation
where higher order terms are saddles with $k$ coinciding giants. The bulk grading  leads to 
cancellations that implement the exclusion principle and are necessary for the index to reproduce black hole physics as shown recently in \cite{Beccaria:2023hip}. \footnote{Outside the context of giant graviton expansions, it is known 
that  states in the index with charge of order $N^{2}$ can reproduce the entropy of supersymmetric black holes in $AdS_{5}$  \cite{Cabo-Bizet:2018ehj,Choi:2018hmj,Benini:2018ywd,Kim:2019yrz,Cabo-Bizet:2019osg,Cabo-Bizet:2019eaf}.}
The extra grading emerging in the giant graviton expansions suggests that open strings on  D-branes are presumably not dual to  physical states of the boundary gauge theory. 
As recently pointed out in  \cite{Lee:2023iil},  they are  objects that  cancel out the redundancies among states  at finite $N$ or, 
in other words, are dual to massive auxiliary ghost states on the boundary. Technically, this holds for normalizable states, while other states associated with open strings on D3 giants are dual to gauge theory 
physical operators with R-charge of order $N$ \cite{Balasubramanian:2002sa,Balasubramanian:2004nb,deMelloKoch:2007rqf}. \footnote{The relation between giant brane configurations and the exclusion principle
was already in the original studies of giant gravitons that were discovered by considering 
large angular momentum BPS particles moving in sphere part of $AdS_{n} \times  S^{m}$. At high momentum they become spherical branes (due to the role of the form fields supporting the 
geometry ) expanding in $S^{m}$ with the stringy exclusion principle being same as condition that brane radius cannot exceed sphere one \cite{McGreevy:2000cw}. 
}
%\footnote{For the so-called dual giants
%ie wrapping an $S^{3}$ inside $AdS_{5}$ and rotating along an equatorial circle of $S^{5}$ \cite{Grisaru:2000zn,Hashimoto:2000zp}
%a similar bound exists for different physical reasons related to charge conservation, see  \cite{Suryanarayana:2004ig,Mandal:2006tk}. }

From the above considerations, a key achievement in this context is the ability to evaluate the brane indices $\I_{k}^{\rm GG}(q)$ in (\ref{1.1})
on the gravity side of any given model. This is interesting already at leading wrapping $k=1$ because the superconformal theory of fluctuations on the BPS wrapped D-branes (breaking 
a fraction of the original supersymmetry) is non-trivial and depends on the wrapping geometry. 

In the approach of Imamura \textit{et al.},  brane indices are obtained by heavily relying on the unbroken superconformal symmetry of the wrapped brane
which is cleverly mapped to the boundary symmetry algebra. This approach provides the brane indices $\I_{k}^{\rm GG}(q)$ in terms of the index of a simpler  boundary SCFT
by analytic continuation in fugacity space.

Recently, the analysis of (leading) finite $N$ corrections to refined superconformal indices was reconsidered  by direct semiclassical analysis of branes wrapped in
a twisted string background with peculiar mixing terms in the metric and supporting forms. The twist terms implement the 
geometrization of  fugacities present in the definition of the index. 

This programme has been accomplished  for the superconformal index of the 6d $(2,0)$ theory on $S^{1}_{\beta}\times S^{5}$ ($\beta$ being the length of the thermal circle)
whose finite $N$ corrections 
are reproduced by semiclassical M2 brane wrapped on $S^{1}\times S^{2}$ in a twisted  M-theory background $AdS_{7}\times S^{4}$ \cite{Beccaria:2023sph}. 
A similar analysis has been performed for the ``mirror'' case of the index of 3d $\N=8$ supersymmetric level-one $U(N)\times U(N)$ ABJM theory 
on $S^{1}_{\beta}\times S^{2}$ whose finite $N$ corrections are computed by 
semiclassical M5 brane wrapped on $S^{1}\times S^{5}$ in twisted $AdS_{4}\times S^{7}$ background \cite{Beccaria:2023cuo}. To recall, the leading non-perturbative
correction to the unrefined index in the two cases had the following form  (see Appendix \ref{app:special} for notation)
\ba
\la{1.3}
\I^{(2,0)}_{N}(q) &= \I^{(2,0)}_{\rm KK}(q) \,\bigg[1-\frac{q}{(1-q)^{2}}\,q^{N}+\mc O(q^{2N})\bigg]\,,
\quad\qquad\qquad\qquad\qquad q = e^{-\beta}, \\
\la{1.4}
\I^{\rm ABJM}_{N}(q') &= \I^{\rm ABJM}_{\rm KK}(q')\, \bigg[1-\frac{1}{6}N^{3}\bigg(\frac{q'}{(q',q')_{\infty}^{7}}+\dots\bigg)\,
q^{N}+\mc O(q'^{2N})\bigg],  \qquad q' = e^{-\frac{1}{2}\beta}\,.
\ea
The correction in (\ref{1.3}) has the form  (\ref{1.1}), with the leading brane index $\I_{k=1}^{\rm GG}(q)$ being a rational function of $q$. The result for ABJM in (\ref{1.4}) is different in two respects. First, 
the dependence on $q$ is no more rational, but involves instead an infinite product. Second, the brane index has an extra $N$ dependence apart from that in the classical
factor $q^{N}$. The correction indeed has an overall $N^{3}$ factor plus subleading corrections in $1/N$ denoted by dots in (\ref{1.4}). This is a common feature of unrefined indices. 
As we mentioned before, the expansion (\ref{1.1}) --
with $N$ independent brane indices $\I_{k}^{\rm GG}(q)$ -- holds at generic points in fugacity space. Additional polynomial dependence on $N$ may  occur
in unrefined indices and, more generally, in refined indices at special varieties defined by algebraic relations between fugacities. This is a form of ``wall-crossing'' phenomenon 
\cite{Gaiotto:2021xce,Lee:2022vig,Beccaria:2023zjw}. As explained in \cite{Beccaria:2023cuo}, in the gravity calculation extra powers of $N$ come from zero modes of fluctuations on the wrapped brane which are 
removed (or regularized) when generic fugacities are switched on.

\subsection{Summary of results}

In this paper,  we apply the same strategy back to the simplest and more familiar case of  4d $\N=4$ $U(N)$ SYM theory where finite $N$ corrections to the index on $S^{1}_{\beta}\times S^{3}$
originate from D3 branes wrapped on $S^{1}\times S^{3}$
in twisted $AdS_{5}\times S^{5}$. In particular, we will focus on  the refined Schur index for which a wealth of useful results are available \cite{Bourdier:2015sga,Bourdier:2015wda,Pan:2021mrw,Beccaria:2023zjw}
and that was treated by the symmetry analytic continuation approach in \cite{Arai:2020qaj}. One motivation for extending the analysis in 
\cite{Beccaria:2023sph,Beccaria:2023cuo} to the $\N=4$ SYM case is that recently a similar (but complementary) 
analysis of the $\frac{1}{2}$-BPS index appeared  in \cite{Eleftheriou:2023jxr}. There, the superconformal index is computed by 
localization of the wrapped brane fluctuations, \ie exploiting supersymmetry from the start. In particular, and by construction, one keeps  
only contributions surviving supersymmetry cancellations. Thus, gapped sectors can be dropped, including non-constant 
modes on $S^3$, 
reducing the problem to the study of a supersymmetric  quantum mechanical 
particle moving in two dimensions in a constant magnetic field.

Instead, our analysis will be more direct and will be based only on the conformal properties of the fluctuations. 
The brane index calculation will thus require a generalization (in presence of various fugacities)
of the operator counting method in \cite{Cardy:1991kr,Kutasov:2000td,Aharony:2003sx}
as in  our previous applications in \cite{Beccaria:2023sph,Beccaria:2023cuo}.
For the unrefined Schur index of $\N=4$ $U(N)$ SYM in $S^{1}_{\beta}\times S^{3}$, 
we will reproduce the expansion of the exact result in \cite{Bourdier:2015wda}, \ie the leading wrapping correction to the index
\be
\la{1.5}
\I^{\rm SYM}_{N}(q) = \I_{\rm KK}(q) \bigg[1-(N+2)\,q^{N+1} +\mc O(q^{2N})\bigg]\,,
\ee
which is  similar to the 6d $(2,0)$ case in (\ref{1.3}). Again, the enhancement factor of $N$ originates from zero modes of fluctuations and is absent 
in the flavored (or refined) Schur index $\I^{\rm SYM}_{N}(q, u)$ depending on an additional fugacity $u$ dual to certain R-charges. In this case, 
finite $N$ corrections have a dependence  on $N$ fully encoded in the classical factor $q^{N}$ with the brane indices being $N$ independent quantities. The detailed generalization of 
 (\ref{1.5}) reads
 \be
\la{1.6}
\I^{\rm SYM}_{N}(q,u) = \I_{\rm KK}(q,u) \bigg[1+\bigg(u^{N+3}\,G(q,u)-u^{-N-1}\,G(q,u^{-1})\bigg)\,\frac{q^{N+1}}{1-u^{2}} +\mc O(q^{2N})\bigg]\,.
\ee
where the function
\be
\la{1.7}
G(q,u) = -\frac{(1-u^{2})^{2}}{u^{2}}\,\frac{(u^{-1}q; u^{-1},q)^{2}_{\infty}}{(u^{-2}, u^{-1}q)_{\infty}(u^{2},u^{-1}q)_{\infty}},
\ee
is smooth  for $u\to 1$ and admits a Taylor expansion around $q=0$.

As a byproduct, the analysis of the index in terms of  a twisted partition function allows one  
to discuss the supersymmetric Casimir energy \cite{Assel:2015nca,Assel:2014tba,Cassani:2014zwa,Bobev:2015kza} of the fluctuations on the wrapped brane.
This quantity $E_{c}^{\rm D3}$ appears in the general relation 
\be
Z_{1}(S_{\beta}^{1}\times S^{3}) = e^{-\beta E_{c}^{\rm D3}}\, \widehat{\I}_{1}(q).
\ee
where $Z_{1}$ is the one-loop partition function of fluctuations and $\widehat{\I}_{1}(q)$ the associated leading brane index.
The resulting value of $E_{c}^{\rm D3}$ will be checked to match the regularized sum over fluctuation modes of the zero point combination of 
Cartan charges defining the index. In particular, for the Schur index, we will get $E_{c}^{\rm D3}(u) = 0$ for generic $u$, and $E_{c}^{\rm D3}(1)=-1$ in the unrefined limit.

To frame our analysis in a more general  perspective, let us  summarize the current status of finite $N$ corrections to superconformal index at large $N$. 
As we overviewed, there are three possible approaches available so far:
\begin{enumerate}
\item[a)] In the first approach, one uses unbroken superconformal symmetry to predict the brane index as the analytic continuation in fugacity space of a simpler boundary SCFT index. 
This was started in  \cite{Arai:2019xmp} by examining the index of S-fold theories and subsequently applied to many other models. 
\item[b)] The second (recent) approach exploits supersymmetry to localize fluctuations on the wrapped brane. In this framework, the analysis is reduced to considering ungapped sectors, 
like the supersymmetric quantum mechanical model of \cite{Eleftheriou:2023jxr} for the $\frac{1}{2}$-BPS index of $\N=4$ SYM.
\item[c)] The  third approach, adopted here and previously in  \cite{Beccaria:2023sph,Beccaria:2023cuo}, directly exploits only conformal symmetry on the brane worldvolume. For each fluctuation, one determines its associated 
partition function on a background with suitable twists related to the index. The different contributions are finally combined in supersymmetric way (fluctuations belong to a multiplet of the preserved superconformal symmetry).
\end{enumerate}
It is certainly useful to have several independent techniques available. However, as we explained, the most interesting physical problems raised by (\ref{1.1}) require going beyond the leading wrapping contribution
in order to analyze the specific cancellations among higher order contributions. It is thus important to critically assess the methods in (a), (b), and (c) from this point of view.
As a general comment, at wrapping order $k$, fluctuations on the wrapped brane are expected to belong to a $U(k)$ gauge theory, living on the stack of $k$ branes \cite{Gaiotto:2021xce}.
In method (a), higher wrapping contributions were computed by introducing \textit{ad hoc} rules to handle  the holonomy integrals of the $U(k)$ theory. 
Finding a general rule seems to be an open problem, see  \cite{Imamura:2022aua,Fujiwara:2023bdc} and \cite{Lee:2022vig}
for recent developments. 
In method (b), $k$-wrapping contributions to the $\frac{1}{2}$-BPS index were (not surprisingly) computed by promoting the supersymmetric quantum mechanical model to a matrix one with fields being $U(k)$ matrices.
This promising approach worked well in the $\frac{1}{2}$-BPS index of $\N=4$ SYM. It remains to see how the localization approach extends to  other models and for refined indices. Finally, method (c) has been applied so far only to the 
leading wrapping contribution to the index. As we mentioned, it has still to be extended to the direct analysis of fluctuations for the multiply wrapped brane by considering suitable conformal $U(k)$ gauge theories for each fluctuation.
%
%
%generalization of  
%This is potentially non-trivial given the known difficulties in reproducing results for multiply wound Wilson loops, see for instance \cite{Bergamin:2015vxa,Forini:2015bgo}. 
We hope that the interplay between the above-mentioned approaches 
will eventually lead to some consistent method to compute at any higher order the finite $N$ contributions to superconformal indices.

\subsection{Plan of the paper}
In Section \ref{sec:index} and \ref{sec:schur} we discuss the algebraic aspects of superconformal index of $\N=4$ theories in 4d and its
Schur specialization for $\N=4$ $U(N)$ SYM. In Section \ref{sec:arai} we summarize the results of \cite{Arai:2020qaj}, \ie
the approach based on exploiting analytic continuation of the preserved superconformal symmetry on the wrapped D3 branes. In Section \ref{sec:twist} we introduce a twist in $AdS_{5}$
and study the fluctuations of D3 branes wrapped on a $S^{1}\times S^{3}$ inside deformed $AdS_{5}\times S^{5}$. The explicit spectrum of scalar, fermionic, and gauge fluctuations
on the brane are determined. All fluctuations turn out to be conformal fields on $S^{1}\times S^{3}$ with constant holonomy along the thermal cycle. By computing the associated partition functions
using an extensions of the operator counting method, we reproduce in Section \ref{sec:index-twist} the leading wrapping finite $N$ contribution to the unrefined index. 
The $N$ dependent enhancement in (\ref{1.5}) is explained in terms of brane fluctuations zero modes, similar to what happens in the ABJM index \cite{Beccaria:2023cuo}.
We also discuss the supersymmetric Casimir energy of the system of fluctuations on the wrapped D3 brane.
Finally, in Section \ref{sec:flavor}, we  deal with the refined index with an extra fugacity coupled to R-charges by adding extra twists in $S^{5}$.

\section{$\N=4$ Superconformal index}
\la{sec:index}

For a $\N=4$ theory on $\mathbb R\times S^{3}$ with $PSU(2,2|4)$ symmetry  the superconformal index   \cite{Kinney:2005ej} is
defined as (notation is $\bm{u} = (u_{1},u_{2},u_{3})$)
\be
\la{2.1}
\I(q,y,\bm{u}) = \Tr_{\rm BPS}[(-1)^{\rm F}q^{H+J+\bar J}y^{2J}u_{1}^{R_{1}}u_{2}^{R_{2}}u_{3}^{R_{3}}]\,, \qquad u_{1}u_{2}u_{3}=1\,,
\ee
where $H$ is the dilatation operator, $J, \bar J$ left and right-handed spins \footnote{They are suitable generators of the   group $SU(2)\times SU(2)$, being the $\mathbb R\times S^{3}$ version of the Lorentz 
group in $\mathbb R^{1,3}$.} 
and 
$R_{I}$ are three Cartan generators of the R-symmetry $\mathfrak{su}(4)$. 
In terms of traceless $R\indices{^{I}_{J}}$ generators of $SU(4)$, with $R\indices{^{I}_{I}}=0$, we have explicitly
$R_{1} = \frac{1}{2}(R\indices{^{1}_{1}}+R\indices{^{2}_{2}}-R\indices{^{3}_{3}}-R\indices{^{4}_{4}})$, 
$R_{2} =\frac{1}{2}(R\indices{^{1}_{1}}-R\indices{^{2}_{2}}+R\indices{^{3}_{3}}-R\indices{^{4}_{4}})$, 
$R_{3} = \frac{1}{2}(R\indices{^{1}_{1}}-R\indices{^{2}_{2}}-R\indices{^{3}_{3}}+R\indices{^{4}_{4}})$.
The BPS condition depends on the specific supercharge used to define the index and can be taken to read 
\be
H-2\bar J-2R\indices{^{1}_{1}}=0\,.
\ee
It is convenient to change basis of fugacities and write
\ba
\I(q,y,u,v) &= \Tr_{\rm BPS}[(-1)^{\rm F}q^{H+J+\bar J}y^{2J}u^{R_{1}-R_{2}}\,v^{R_{2}-R_{3}}]\,, \\
\la{2.4}
u &= u_{1}, \qquad v = u_{1}u_{2}\,.
\ea
For an arbitrary $\N = 2$ superconformal theory there is a corresponding chiral algebra 
and the Schur index \cite{Gadde:2011uv} is a specialization of the superconformal index giving the chiral algebra vacuum character \cite{Beem:2013sza}.
In our conventions, the Schur limit is $y=v=1$  \footnote{Often, the index is defined tracing $q^{H+\bar J}\bar y^{2J}$ and Schur limit is $\bar y = q^{\frac{1}{2}}$. 
In our notation $\bar y = q^{\frac{1}{2}}y$
and the Schur limit  corresponds to  $y=1$. The Schur index gets contributions only from states obeying 
the additional BPS condition $H+2J+2R\indices{^{4}_{4}}=0$ and 
this allows to replace $H+J+\bar J = H-R_{3}$. }, \ie the trace 
\be
\la{2.5}
\I(q,u) = \Tr_{\rm BPS}[(-1)^{\rm F}q^{H+J+\bar J}u^{R_{1}-R_{2}}]\,.
\ee

\subsection{The $\N=4$ Maxwell multiplet}

The $\N=4$ Maxwell multiplet is a single free vector multiplet with components $F_{ab}, \l_{Ia}, \phi_{IJ}, \bar\l^{I\dot a}, \bar F^{\dot a\dot b}$,
where $I,J$ are $SU(4)_{R}$ indices and $\phi_{IJ}$ is in the 2-antisymmetric representation (with a reality condition). This multiplet transforms as  
a short representation of the superconformal algebra and its index is obtained from the plethystic exponential  
\be
\I^{\rm Maxwell}(q,y,u,v) = \PE[\Isp^{\rm Maxwell}(q,y,u,v)] = \exp\bigg[\sum_{n=1}^{\infty}\frac{1}{n}\Isp^{\rm Maxwell}(q^{n},y^{n},u^{n},v^{n})\bigg]\,,
\ee
with the single particle index \cite{Dolan:2002zh}
\be
\la{2.7}
\Isp^{\rm Maxwell}(q,y,u,v) = \frac{q\,(u+vu^{-1}+v^{-1})-(q^{2}y+qy^{-1})-q^{2}(u^{-1}+uv^{-1}+v)+2q^{3}}{(1-q^{2}y)(1-q y^{-1})}\,.
\ee
In the unflavored Schur limit $y=u=v=1$, we get 
\be
\Isp^{\rm Maxwell}(q) \equiv \Isp^{\rm Maxwell}(q,1,1,1) = \frac{2q-2q^{2}}{1-q^{2}} = \frac{2q}{1+q}\,,
\ee
where $-2q^{2}$ comes from the unflavored indices of a $\N=2$ vector multiplet and $2q$ from a  $\N=2$ hypermultiplet.

\subsection{Schur index of $\N=4$ $U(N)$ SYM}
\la{sec:schur}
 
For a gauge theory with multiplets in the   representations $\mc R$ of the gauge group, the index is \cite{Romelsberger:2005eg,Kinney:2005ej}
\be
\la{2.9}
\I_{N}(q,y,u,v) = \int \mc DU\prod_{\mc R}\exp\bigg[\sum_{n=1}^{\infty}\frac{1}{n}\Isp^{\mc R}(q^{n},y^{n},u^{n},v^{n})\, \chi_{\mc R}(U^{n})\bigg]\,,
\ee
where integration is over Haar measure of gauge group and $\chi$ is a representation character. 
For $\N=4$ SYM with gauge group $U(N)$, we need to consider a Maxwell multiplet  in the adjoint of $U(N)$. In the unflavored case, this 
gives the expression \cite{Bourdier:2015wda} 
\be
\la{2.10}
\I^{\rm SYM}_{N}(q) = \frac{q^{-N^{2}/4}\eta(\tau)^{3N}}{\pi^{N}N!}\int_{0}^{\pi}d^{N}\bm{\alpha}
\frac{\prod_{i<j}\vartheta_{1}(\alpha_{i}-\alpha_{j},q)^{2}}{\prod_{i,j}\vartheta_{4}(\alpha_{i}-\alpha_{j},q)}\,,
\ee
where $\bm{\alpha}$  are gauge group holonomies and definitions of elliptic functions are  in  Appendix \ref{app:special}.

The index may be evaluated at finite $N$ expanding in powers of $q$ and integrating over  $\bm\alpha$. For the lowest values of $N$, this gives
\bea
\la{2.11}
\I^{\rm SYM}_{1}(q) &= 1+2q+q^{2}+2q^{3}+2q^{4}+\cdots, \\
\I^{\rm SYM}_{2}(q) &= 1+2q+4q^{2}+4q^{3}+6q^{4}+\cdots, \\
\I^{\rm SYM}_{3}(q) &= 1+2q+4q^{2}+8q^{3}+9q^{4}+\cdots,  \\
\I^{\rm SYM}_{4}(q) &= 1+2q+4q^{2}+8q^{3}+14q^{4}+\cdots.
\eea
The exact result was derived in \cite{Bourdier:2015wda} from (\ref{2.10}) and takes the giant graviton expansion form 
\be
\la{2.12}
\I^{\rm SYM}_{N}(q) = \frac{1}{\vartheta_{4}(0)}\sum_{k=0}^{\infty}(-1)^{k}\bigg[\binom{N+k}{N}+\binom{N+k-1}{N}\bigg]\,q^{kN+k^{2}}\,.
\ee
The coefficients of the expansions (\ref{2.11}) stabilize for $N\to \infty$ and correspond to the series
\be
\la{2.13}
\I_{\rm KK}(q) \equiv \I^{\rm SYM}_{\infty}(q) = \frac{1}{\vartheta_{4}(0)} = \prod_{n=1}^{\infty}(1-q^{2n})^{-1}(1-q^{2n-1})^{-2} = \prod_{n=1}^{\infty}\frac{1+q^{n}}{1-q^{n}}\,,
\ee
whose explicit first terms are 
\be
\I_{\rm KK}(q) = 1+2q+4q^{2}+8q^{3}+14q^{4}+18q^{5}+28 q^{6}+\cdots.
\ee
The generating function $\I_{\rm KK}(q)$ reproduces the index of  BPS Kaluza-Klein modes of the supergravity multiplet in $AdS_{5}\times S^{5}$ \cite{Kinney:2005ej}.
Including systematically the large $N$ exponentially suppressed terms  in (\ref{2.12}), we can write the  large $N$ expansion of the unflavored index as
\be
\la{2.15}
\I^{\rm SYM}_{N}(q) = \I_{\rm KK}(q) \bigg[1-(N+2)\,q^{N+1} +\frac{1}{2}(N+1)(N+4)\,q^{2N+4}+\mc O(q^{3N})\bigg]\,.
\ee
The flavored Schur index 
 is obtained starting from the following replacement in the Maxwell single particle index
 \be
 \la{2.16}
\Isp^{\rm Maxwell}(q) = \frac{2q-2q^{2}}{1-q^{2}}\quad\to\quad \Isp^{\rm Maxwell}(q,u)  \equiv \Isp^{\rm Maxwell}(q,1,u,1)= \frac{q(u+u^{-1})-2q^{2}}{1-q^{2}}\,.
\ee
The first term in the numerator comes from the $\N=2$ hypermultiplet and is due to the extra fugacity $u^{R'}$ where $R'$ is a R-symmetry Cartan taking values $\pm 1$ 
on the complex scalar and its conjugate in that multiplet.
In this case, the large $N$ Kaluza-Klein factor takes the form \footnote{
For $u=1$, we get 
$\I_{\rm KK}(q, 1) = \PE\bigg[\frac{2q}{1-q}-\frac{q^{2}}{1-q^{2}}\bigg] = \PE\bigg[\frac{1}{2}\sum_{k=1}^{\infty}(3-(-1)^{k})q^{k}\bigg] = 
\prod_{k=1}^{\infty}\left(\frac{1}{1-q^{k}}\right)^{\frac{1}{2}(3-(-1)^{k})}
= (1-q)^{-2}(1-q^{2})^{-1}(1-q^{3})^{-2}(1-q^{4})^{-1}(1-q^{5})^{-2}(1-q^{6})^{-1}\cdots$, 
in agreement with (\ref{2.13}).
}
\be
\I_{\rm KK}(q, u) = \PE\bigg[\frac{uq}{1-uq}+\frac{u^{-1}q}{1-u^{-1}q}-\frac{q^{2}}{1-q^{2}}\bigg]\,.
\ee
For $u\neq 1$, we don't have  a simple  explicit expression for the index $\I_{N}^{\rm SYM}(q,u)$ at finite $N$ generalizing (\ref{2.12})
and a derivation of the large $N$ expansion (\ref{2.15}) is non-trivial starting from the localization formula (\ref{2.9}).

\section{ D3-brane index from analytic continuation}
\la{sec:arai}

In the approach of Imamura \textit{et al.}, see in particular \cite{Arai:2019xmp,Imamura:2021ytr}, the general index (\ref{2.1}) for $\N=4$ $U(N)$ SYM can be computed at large $N$ 
as the triple sum
\be
\la{3.1}
\I_{N}^{\rm SYM}(q,y,\bm{u}) = \I_{\rm KK}(q,y,\bm{u})\,\sum_{\bm{k}=\bm{0}}^{\infty}(qu_{1})^{k_{1}N}\,(qu_{2})^{k_{2}N}\,(qu_{3})^{k_{3}N}\,\I^{\rm D3}_{\bm{k}}(q,y,\bm{u})\,, 
\qquad \bm{k} = (k_{1},k_{2},k_{3})\,.
\ee
Notice the slight change of notation $\widehat{\I}_{k}\to \I^{\rm D3}_{\bm k}$, \cf (\ref{1.1}), to emphasize the specific case under consideration.
The $N$-independent factor $\I_{\rm KK}(q,y,\bm{u})$ is the $N\to\infty$ index, {\em i.e.} the index counting gauge invariant operators at large-$N$, ignoring trace relations, 
and agrees with the index from BPS Kaluza-Klein supergravity modes on $AdS_{5}\times S^{5}$.
The triple sum in (\ref{3.1}) represents a sum  over D3 branes wrapped $k_{I}$ times, $I=1,2,3$,  over  the cycle $\sZ_{I}=0$ in $S^{5}$, \ie  $|\sZ_{1}|^{2}+|\sZ_{2}|^{2}+|\sZ_{3}|^{2}=1$.
The generators $R_{I}$ in (\ref{2.1}) are associated with phase rotations in the $\sZ_{I}$ plane. 
The classification of the relevant BPS wrapped branes was analyzed in \cite{Mikhailov:2000ya,Mikhailov:2002wx}
where it was shown how to build  supersymmetric cycles in backgrounds $AdS_{n}\times S^{m}$ that generalize giant gravitons. 
In particular, for D3 branes in $AdS_{5}\times S^{5}$, any analytic constraint $f(\sZ_{1}, \sZ_{2}, \sZ_{3})=0$ defines a supersymmetric configurations which is 
$\frac{1}{2}$, $\frac{1}{4}$, or $\frac{1}{8}$-BPS depending on the number of arguments of $f$. The simplest case $f(\sZ_{1})=0$ gives a $\frac{1}{2}$-BPS solution and multiple zeroes are associated with a number of 
concentrical spherical giants,  see \cite{Abbott:2013ija} for a discussion of the related geometries. \footnote{
The  sum over three cycles $\sZ_{I}=0$ in (\ref{3.1}) deserves a comment. Looking for simplicity at wrapping one,  we need to consider the family of configurations $a\sX+b\sY+c\sZ=0$. It is
 parametrized by $\mathbb{CP}^{2}$ and we need to sum over fluctuations around a representative in each of the three covering patches $\sZ_{I}\neq 0$.
}

The factor $\prod_{I}(qu_{I})^{k_{I}N}$ in (\ref{3.1}) comes from the classical 
action of the wrapped brane. Finally, the index $\I^{\rm D3}_{\bm{k}}$ refers to the gauge theory on the wrapped D3 brane which is a $G_{\bm k} = U(k_{1})\times U(k_{2})\times U(k_{3})$
quiver gauge theory with three bifundamental hypermultiplets. This index is by construction independent on $N$. Thus, the expansion (\ref{3.1}) has the form of a giant graviton expansion
with each term having the weight $q^{kN+\delta_{k}}$, where $k=k_{1}+k_{2}+k_{3}$ is the total wrapping and $\delta_{k}$ is a non-negative integer.

As conjectured in \cite{Imamura:2021ytr}, the brane index  $\I^{\rm D3}_{\bm k}$  is given by a $G_{\bm k}$ holonomy integral similar in structure to (\ref{2.9}),
\ie schematically, 
\be
\la{3.2}
\I^{\rm D3}_{\bm k}(q,y,\bm{u}) = \int \mc D{\bm U}\ \PE[\I^{{\rm D3}, {\rm vec}}_{\bm k}(q,y,\bm{u})\,\chi^{\rm vec}(\bm U)]\,
\PE[\I^{{\rm D3}, {\rm hyp}}_{\bm k}(q,y,\bm{u})\,\chi^{\rm hyp}(\bm{U})]\,,
\ee
where  $\I^{{\rm D3}, {\rm vec}}_{\bm k}$, $\I^{{\rm D3}, {\rm hyp}}_{\bm k}$ are the contributions from vector and hypermultiplets in the quiver gauge theory.
In (\ref{3.2}), integration over $\bm{U}= (U_{1},U_{2},U_{3})$ with $U_{I}\in U(k_{I})$ is a schematic way to represent the integration 
over holonomies that implements singlet projection. Similarly, the factors $\chi^{\rm vec}(\bm{U})$ and $\chi^{\rm hyp}(\bm U)$ represent suitable group characters, \cf (\ref{2.9}). 
Details of the integration cycles are a key issue not completely understood and recently discussed in \cite{Imamura:2022aua}. Here, we will focus on total wrapping $k=1$
where holonomies are trivial and also quiver hypermultiplets are absent. Hence, after the relabeling $(\sZ_{1}, \sZ_{2}, \sZ_{3}) \to (\sX, \sY, \sZ)$ for simpler notation, the expansion (\ref{3.1}) reads 
\be
\la{3.3}
\I_{N}^{\rm SYM}(q,y,\bm{u}) = \I_{\rm KK}\bigg[1+q^{N}\bigg(u_{1}\, \I^{\rm D3}_{\sX}(q,y,\bm{u})+
u_{2}\, \I^{\rm D3}_{\sY}(q,y,\bm{u})+u_{3}\, \I^{\rm D3}_{\sZ}(q,y,\bm{u})\bigg)+\mc O(q^{2N})\bigg]\,,
\ee
where $\I^{\rm D3}_{\sX} \equiv \I^{\rm D3}_{1,0,0} = \I^{{\rm D3}, {\rm vector}}_{1,0,0}$, \textit{etc.} The vector multiplet 
index on D3 is again the plethystic exponential of a single particle index conjectured in \cite{Arai:2019xmp,Imamura:2021ytr} to be  
\footnote{The definition of the index in \cite{Imamura:2021ytr} has fugacities $q^{H+\bar J}\bar y^{2J} u_{1}^{R_{1}} u_{2}^{R_{2}} u_{3}^{R_{3}}$.
To match our definition of the index we need to identify $\bar y = q^{\frac{1}{2}}y$.}
\be
\la{3.4}
\Isp^{\rm D3}_{\sX}(q,y,\bm{u}) = 1-\frac{(1-q^{-1} u_{1}^{-1})(1-q^{2}y)(1-qy^{-1})}{(1-q u_{2})(1-q u_{3})}\,,
\ee
Using (\ref{2.4}), this reads, \cf Eq.~(4.35) of \cite{Arai:2019xmp}, 
\ba
\la{3.5}
\Isp^{\rm D3}_{\sX}(q,u,v) &= \frac{\frac{1}{qu}-q(y+\frac{1}{qy})\frac{1}{u}-q(\frac{1}{v}+\frac{v}{u})+q^{2}(y+\frac{1}{qy})+q^{2}\frac{2}{u}-q^{3}}{(1-q/v)(1-q v/u)}\,.
\ea
The expressions for $\Isp^{\rm D3}_{\sY}$, $\Isp^{\rm D3}_{\sZ}$ are similarly obtained from (\ref{3.4}) by permuting $u_{1},u_{2},u_{3}$.

The index (\ref{3.5}) is computed in two ways, both relying on the broken superconformal symmetry on the D3 brane.
By a direct computation, see Appendix A of \cite{Imamura:2021ytr}, the modes on the brane are organized in multiplets of the reduced superalgebra (the wrapped brane breaks half of the 32 supersymmetries
and preserves a  $SU(2|2)\times SU(2|2)$ superalgebra)
and the index is evaluated by summing the separate contributions with the result (\ref{3.5}). Alternatively, one can rely on the automorphism between the supersymmetry algebra on the boundary and  on the wrapped D3-brane.
This gives the analytic continuation relation
\bea
\la{3.6}
& \Isp^{\rm D3}_{\sX}(q,y,u,v) = \Isp^{\rm Maxwell}(q',y',u',v')\,,  \\
& q' = q^{\frac{2}{3}}u^{-\frac{1}{3}}, \quad
y' = q^{-\frac{1}{3}}u^{-\frac{1}{3}}v, \quad
u' = q^{-\frac{5}{3}}u^{-\frac{2}{3}}, \quad
v' = q^{-\frac{1}{3}}y u^{-\frac{1}{3}}\,, 
\eea
that expresses the wrapped brane index in terms of the index of the Maxwell multiplet living on the boundary.  Starting from (\ref{2.7})
and using (\ref{3.6}), one gets again (\ref{3.5}). This second method is somehow impressive since it reduces the wrapped brane index to 
a computation in the free boundary conformal theory. Similar analytic continuations have been exploited 
in later works on finite $N$ correction to the index in different models, like the M-brane indices considered in \cite{Arai:2020uwd}.

\subsection{Schur specialization}

The refined Schur index is obtained by specializing $y=v=1$. From (\ref{3.6}) we get 
\be
\Isp^{\rm D3}_{\sX}(q,u) = \Isp^{\rm Maxwell}( q^{\frac{2}{3}}u^{-\frac{1}{3}},q^{-\frac{1}{3}}u^{-\frac{1}{3}}, q^{-\frac{5}{3}}u^{-\frac{2}{3}},q^{-\frac{1}{3}} u^{-\frac{1}{3}})\,.
\ee
For $y=v$ the following scaling invariance holds
\be
\Isp^{\rm Maxwell}(q,y,u,y) = \Isp^{\rm Maxwell}(\l^{-1}q,\l^{2}y,\l u,\l^{2}y)\,.
\ee
Hence, taking $\l=(q^{-\frac{1}{3}}u^{-\frac{1}{3}})^{-\frac{1}{2}} = (qu)^{\frac{1}{6}}$ we can write $\Isp^{\rm D3}_{\sX}(q,u)$ in terms of the Maxwell
Schur index
\be
\la{3.9}
\Isp^{\rm D3}_{\sX}(q,u) = \Isp^{\rm Maxwell}( (qu)^{-\frac{1}{6}}q^{\frac{2}{3}}u^{-\frac{1}{3}},(qu)^{\frac{1}{6}} q^{-\frac{5}{3}}u^{-\frac{2}{3}})
 = 
\Isp^{\rm Maxwell}(q^{\frac{1}{2}}u^{-\frac{1}{2}}, q^{-\frac{3}{2}}u^{-\frac{1}{2}})\,.
 \ee
Using (\ref{2.7}), or taking the Schur limit in (\ref{3.5}), this is 
\be
\la{3.10}
\Isp^{\rm D3}_{\sX}(q,u) =  \frac{\frac{1}{uq}-\frac{2}{u}q+q^{2}}{1-\frac{1}{u}q}\,.
\ee
As discussed in \cite{Arai:2020qaj}, the general formula (\ref{3.1}) simplifies to 
\be
\la{3.11}
\I_{N}^{\rm SYM}(q,u) = \I_{\rm KK}(q,u)\, \sum_{k_{1},k_{2}=0}^{\infty}
 (uq)^{k_{1}N}\,(u^{-1}q)^{k_{2}N}\,  q^{2k_{1}k_{2}}\    \I_{\sX, k_{1}}^{\rm D3}(q,u)\, \,\I_{\sY, k_{2}}^{\rm D3}(q,u)\,.
 \ee
 In (\ref{3.11}), we recognize the index of a system  of $k_{1}$ D3 branes and $k_{2}$ D3 branes wrapped on the two  
3-cycles $\sX=0$ and $\sY=0$. Also, due to the Weyl reflection relation 
\be
\la{3.12}
\I_{\sY,k}^{\rm D3}(q,u)=\I_{\sX,k}^{\rm D3}(q,u^{-1})\,,
\ee
one can compute (\ref{3.11}) from the $\sX=0$ index $\I_{\sX,k}^{\rm D3}$ alone. As in (\ref{3.9}), this is  related by analytic continuation to the 
Schur index of  the $\N=4$ SYM theory with gauge group $U(k)$
\ba
\la{3.13}
\I_{\sX, k}^{\rm D3}(q,u) &= \I_{k}^{\rm SYM}(q^{\frac{1}{2}}u^{-\frac{1}{2}}, q^{-\frac{3}{2}}u^{-\frac{1}{2}})\,.
\ea
In particular, the leading contribution to the giant-brane expansion is 
\be
\la{3.14}
\I_{N}^{\rm SYM}(q,u) = \I_{\rm KK}(q,u)\, \bigg[1+q^{N}\bigg(
u^{N}\, \I_{\sX}^{\rm D3}(q,u)+u^{-N}\,\I_{\sX}^{\rm D3}(q,u^{-1})\bigg)+\mc O(q^{2N})\bigg]\,,
 \ee
 where
 \be
 \I_{\sX}^{\rm D3}(q,u) \equiv \I_{\sX, 1}^{\rm D3}(q,u) = \PE[\Isp_{\sX}^{\rm D3}(q,u)]\,.
 \ee
 Using  expression (\ref{3.10}), this evaluates to 
 \be
\I_{\sX}^{\rm D3}(q,u) = -\frac{u q}{1-u q}\prod_{n=1}^{\infty}\frac{[1-(u^{-1}q)^{n}]^{2}}{[1-u^{-2}(u^{-1}q)^{n-1}][1-u^{2}(u^{-1}q)^{n+1}]}\,,
\ee
or, in terms of q-Pochhammer functions (\ref{A.3}),
\be
\I_{\sX}^{\rm D3}(q,u) = -uq\,(1-u^{2})\, \frac{(u^{-1}q,u^{-1}q)_{\infty}^{2}}{(u^{-2},u^{-1}q)_{\infty}\,(u^{2},u^{-1}q)_{\infty}}\,.
\ee
This expression is exact and its expansion in powers of $q$ gives 
\be
\la{3.18}
\I_{\sX}^{\rm D3}(q,u) = \frac{u^{3}}{1-u^{2}}\,q+(1-u^{2})q^{2}+\bigg(\frac{1}{u^{3}}-u^{3}\bigg)\, q^{3}
+\bigg(\frac{1}{u^{6}}-\frac{1}{u^{2}}+1-u^{4}\bigg)\, q^{4}+\cdots\,.
\ee
Notice that one cannot use relation (\ref{3.13}) to obtain  an expansion like (\ref{3.18}) for $k=1$, unless $\I_{k}(q,u)$ is known in exact form and not as a series in $q$.
This is because of negative powers of $q$ in (\ref{3.13}), requiring resummation if the latter identity is naively applied to 
a series in $q$. \footnote{To explain, if we have $f(q) = \frac{1}{1-q}$ which has regular expansion in $q$, 
we can apply $q\to 1/q$ and obtain $f(q^{-1}) =-\frac{q}{1-q}$ which also has a regular expansion in $q$. Of course, this is not visible (in particular in more complicated cases) if we 
apply $q\to 1/q$ to $1+q+q^{2}+q^{3}+\cdots$.  
}
For this reason, in  \cite{Arai:2020qaj}, the expansion (\ref{3.18}) (and its generalization to $n>1$ ) 
was obtained  by applying the analytic continuation relation (\ref{3.13}) \underline{inside} 
the gauge holonomy integral expression of $\I_{k}^{\rm SYM}$. After that, the integration over holonomies was evaluated by a conjectured { ad hoc choice of the contributing poles}.
This  is a major difficulty for next-to-leading (\ie higher wrapping) contributions that we do not discuss here. Nevertheless, in Appendix \ref{app:w2}, we present a direct application of 
analytic continuation for the~$k=2$ brane index starting from the exact expression of the SYM index and avoiding pole selection rules at all.

\subsection{Unrefined index and $u\to 1$ limit}

Let us show how the (leading) giant graviton expansion of the unflavored index in (\ref{2.15}) is reproduced in the $u\to 1$ limit. This comes from a non-trivial combination of the two
terms in (\ref{3.14}) which are separately singular for $u\to 1$, but with a smooth sum.

We  isolate the singular $u\to 1$ contribution to $\I_{\sX}^{\rm D3}(q,u)$ by writing
\ba
\I_{\sX}^{\rm D3}(q,u) &=  \frac{u^{3}}{1-u^{2}}\,q\,G(q,u), \qquad G(q,u) = \frac{1}{1-u q}\prod_{n=1}^{\infty}\frac{[1-(u^{-1}q)^{n}]^{2}}{[1-u^{-2}(u^{-1}q)^{n}][1-u^{2}(u^{-1}q)^{n+1}]},
\ea
where $G(q,u)$ is smooth for $u\to 1$. The value at $u=1$ is 
\be
G(q,1) = \frac{1}{1-q}\prod_{n=1}^{\infty}\frac{1-q^{n}}{1-q^{n+1}} = 1.
\ee
Also, we can compute  the derivative with respect to $u$ at $u=1$ as follows. Taking logarithm, we get 
\ba
\log G(q,u) &= -\log(1-uq)+\sum_{n=1}^{\infty}\bigg[2\log(1-(q/u)^{n})-\log(1-u^{-2}(q/u)^{n})-\log(1-u^{2}(q/u)^{n+1})\bigg]\,,
\ea
and differentiation with respect to $u$ gives after some rearrangement
%\ba
%\frac{\partial_{u}G(q,1)}{G(q,1)} = \frac{q}{1-q}+\sum_{k=1}^{\infty}\bigg[(k-1)\bigg(\frac{1}{1-q^{k}}-\frac{1}{1-q^{k+1}}\bigg)-\frac{q^{k}}{1-q^{k}}\bigg]
%\ea
%\ba
%\frac{G'(q,1)}{G(q,1)} = \frac{q}{1-q}+\sum_{k=1}^{\infty}\bigg[(k-1)\bigg(\frac{1}{1-q^{k}}-1+1-\frac{1}{1-q^{k+1}}\bigg)-\frac{q^{k}}{1-q^{k}}\bigg]
%\ea
%\ba
%\frac{G'(q,1)}{G(q,1)} = \frac{q}{1-q}+\sum_{k=1}^{\infty}\bigg[(k-1)\bigg(\frac{q^{k}}{1-q^{k}}-\frac{q^{k+1}}{1-q^{k+1}}\bigg)-\frac{q^{k}}{1-q^{k}}\bigg]
%\ea
\ba
\frac{\partial_{u}G(q,1)}{G(q,1)} = \frac{q}{1-q}+\sum_{n=1}^{\infty}\bigg[(n-2)\frac{q^{n}}{1-q^{n}}-(n-1)\frac{q^{n+1}}{1-q^{n+1}}\bigg] = 0, \quad\to\quad \partial_{u}G(q,1) = 0\,.
\ea
Hence, 
\bea
& u^{N}\, \I_{\sX}^{\rm D3}(q,u)+u^{-N}\,\I_{\sX}^{\rm D3}(q,u^{-1}) = \bigg[\frac{u^{N+3}}{1-u^{2}} G(q,u) +  \frac{u^{-N-3}}{1-u^{-2}} G(q,-u)\bigg]\, q  \\
& = -[G(q,1)(N+2)+\partial_{u}G(q,1)]\,q+\mc O(u-1) = -q(N+2)+\mc O(u-1).
\eea
reproducing the enhancement in (\ref{2.15}).

\section{D3-brane in twisted $AdS_{5}\times S^{5}$ and unflavored index}
\la{sec:twist}

As we have seen, the large $N$  non-perturbative corrections to the Schur index are expressed in terms of the single quantity $\Isp^{\rm D3}_{\sX}(q,u)$ in (\ref{3.10})
by using (\ref{3.11}) and (\ref{3.12}). Here, we want to reproduce $\Isp^{\rm D3}_{\sX}(q,u)$ by a genuine one-loop computation 
of the partition function on the wrapped D3 brane with a twisted metric that geometrizes the $J,\bar J$ and Cartan charge terms in (\ref{2.5}).
A similar approach was used in \cite{Beccaria:2023sph,Beccaria:2023cuo} as we reviewed in the Introduction.

Starting from the unflavored case $u=1$, the aim is thus to reproduce the remarkably simple expression \footnote{Simplicity of (\ref{4.1}), \ie reduction to a finite number of modes on the wrapped brane, suggests  a treatment by localization as in \cite{Eleftheriou:2023jxr}.}
\be
\la{4.1}
\Isp^{\rm D3}_{\sX}(q)\equiv \Isp^{\rm D3}_{\sX}(q,1) = \frac{\frac{1}{q}-2q+q^{2}}{1-q} = \frac{1}{q}+1-q\,,
\ee
by considering the semiclassical expansion of the D3 brane action on the background (scaling away  common radius) 
\footnote{Here, the coordinate $y$ should not be confused with the fugacity in the index definition.}
\bea
ds^{2} &= ds^{2}_{\widetilde{AdS}_{5}}+ds^{2}_{S^{5}}\,, \\
ds^{2}_{\widetilde{AdS}_{5}} &= dx^{2}+\sinh^{2}x\, d\widetilde{S}_{3}+\cosh^{2}x\, d\tau^{2}\,,  \qquad \tau=\tau+\beta\,, \\
d\widetilde{S}_{3} &= d\psi_{1}^{2}+\sin^{2}\psi_{1}\,d\psi_{2}^{2}+\sin^{2}\psi_{1}\sin^{2}\psi_{2}\,(d\psi_{3}+i\kappa d\tau)^{2}\,,\\
ds^{2}_{ S^{5}} &= dw^{2}+\cos^{2}w\, dS_{3}+\sin^{2}w\,dz^{2}\,.
\eea
The mixing between $\psi_{3}$ and $\tau$ corresponds to the Cartan combination $H+J+\bar J$ in (\ref{2.4}).
The physical (supersymmetric) value of the parameter $\kappa$ is $\kappa=1$, but it will be useful to keep it as a free parameter for the time being.

In a static gauge, the D3 coordinates are $\xi=(\xi^{1}, \dots, \xi^{4}) \equiv (\tau, p\in S^{3})$, \ie the brane wraps $S^{1}_{\beta}\times S^{3}$ where $S^{1}_{\beta}$ is in $\widetilde{AdS}_{5}$ and has length $\beta$,
while $S^{3}\subset S^{5}$. The fluctuating coordinates are $(x, \psi_{i},w,z)$ with background values $x=w=z=0$. These transverse fluctuations are  
4 in $AdS_{5}$ and 2 in $S^{5}$
for a total of 6 scalar fluctuations. To these we will add fermions $\theta$ and the gauge field $A$ on the brane. At semiclassical level, the partition function is \footnote{
Recall that $S_{\rm cl}$ has an explicit overall factor of the D3 brane tension $\T_{3}$.}
\ba
\la{4.3}
Z &= \int \mc DX\, \mc DA\, \mc D \theta\, e^{-S[X, A, \theta]} = Z_{1}\, e^{-S_{\rm cl}}\,(1+\mc O(\T_{3}^{-1}))\,, \\
\la{4.4}
Z_{1} &= e^{-F}, \qquad F = \frac{1}{2}\sum_{a}(-1)^{{\rm F}_{a}}\log\det\Delta_{a}\,,
\ea
where $a$ labels the fluctuation fields (bosonic and fermionic) and $\Delta_{a} = -\nabla^{2}+\cdots$ are the associated 2nd order operators for quadratic fluctuations.
They are defined on $S^{1}_{\beta}\times S^{3}$ with the induced metric on the brane worldvolume. All fields (including fermions) are periodic on the ``thermal'' circle $S^{1}_{\beta}$
for consistency with  supersymmetry.

\subsection{Classical D3 action}

The classical D3 brane action $S_{\rm cl}$ gives the (unrefined) overall factor $q^{N}$ in (\ref{3.3}) and its evaluation is straightforward. 
After a suitable scaling of the gauge field, the bosonic part of the 
supersymmetric D3 brane action is the sum of the Born-Infeld and the Wess-Zumino actions \cite{Bergshoeff:1996tu} (we have no background antisymmetric field $B_{ab}$ here). In Euclidean notation, 
\be
S_{\rm DBI} = \T_{3}\int d^{4}\xi\, \sqrt{\det(G_{ab}+2\pi F_{ab})}, \qquad
S_{\rm WZ} = -\int C_{4}\,,
\ee
with D3 tension
\be
\T_{3} = \frac{N}{2\pi^{2}}\,.
\ee
The 4-form on $S^{5}$ is 
%\footnote{This has the  
% correct normalization  back in Minkowski space
%$\int_{S^{5}}dC_{4} = - \frac{2N}{\pi^{2}}\int_{S^{5}}\cos^{3} w\sin w \,dw\wedge dz\wedge \text{vol}_{S^{3}}
%= \frac{2N}{\pi^{2}}\text{Vol}(S^{5}) = 2\pi N$.
%} 
\be
\la{5.3}
C_{4} = i\T_{3}\,\cos^{4}w\, dz\wedge \text{vol}_{S^{3}}\,.
\ee
The classical action is thus the sum of 
\ba
S_{\rm cl, DBI} &= \T_{3}\times \beta\times  \text{Vol}(S^{3}) = \frac{N}{2\pi^{2}}\beta (2\pi^{2}) = N \beta\,, \qquad
S_{\rm cl, WZ} = 0\,.
\ea
As expected, the total $S_{\rm cl} = N\beta$ corresponds to the factor $q^{N}$ in (\ref{3.3}), where as usual we denote $q=e^{-\beta}$.
%
%
%\separator
%{\sf Arai coordinates}
%
%\noindent
%Untwisted Background (scaling away the common radius)
%\ba
%ds^{2} =  & -\cosh^{2}r\, dt^{2}+dr^{2}+\sinh^{2}r\, dS_{3}  \lp
%+d\theta^{2}+\sin^{2}\theta d\phi^{2}+\cos^{2}\theta[d\psi_{1}^{2}+\sin^{2}\psi_{1}(
%d\psi_{2}^{2}+\sin^{2}\psi_{2}d\psi_{3}^{2})].
%\ea 
%D3 coordinates $(t, \psi_{1},\psi_{2},\psi_{3})$. Fluctuating coordinates $(r, S^{3}, \theta, \phi)$ with background values $r=0$, 
%arbitrary point in $S^{3}\subset AdS_{5}$, $\theta=0$, arbitrary $\phi$.
%
%\noindent
%For the D3 brane without $B_{ab}$ background field and with a suitable rescaling of the gauge field
%\be
%S_{\rm DBI} = -{\rm T}_{3}\int d^{4}\xi\, \sqrt{-\det(G_{ab}+2\pi F_{ab})}, \qquad
%S_{\rm CS} = \int C_{4},
%\ee
%with 
%\be
%{\rm T}_{3} = \frac{N}{2\pi}.
%\ee
%The 4-form is 
%\be
%C_{4} = \frac{N}{2\pi^{2}}\sqrt{g_{S^{3}}}\cos^{4}\theta\ d\phi\wedge d\psi_{1}\wedge d\psi_{2}\wedge d\psi_{3}, \qquad \sqrt{g_{S^{3}}} = \sin^{2}\psi_{1}\sin\psi_{2}
%\ee
%with the correct normalization
%\ba
%\int_{S^{5}}dC_{4} &= - \frac{2N}{\pi^{2}}\int_{S^{5}}\sqrt{g_{S^{3}}}\cos^{3}\theta\sin\theta d\theta\wedge d\phi\wedge d\psi_{1}\wedge d\psi_{2}\wedge d\psi_{3}\lp
%= \frac{2N}{\pi^{2}}\text{Vol}(S^{5}) = 2\pi N.
%\ea

\subsection{Wrapped brane fluctuation one-loop determinant and brane index}

For each separate fluctuation, the free energy in (\ref{4.4})
splits into the "Casimir" part linear in $\beta$ and the genuine thermal part which is exponentially suppressed at large $\beta$, see \eg \cite{Gibbons:2006ij,Giombi:2014yra},
\be
F(\beta) = \beta\, E_{c}+\bar{F}(\beta)\,,
\ee
where $E_{c}$ is the Casimir contribution, the dominant term at low temperature $\beta\to +\infty$.
Denoting by $E_{n}$ the eigenvalues of the fluctuation operator on $S^{3}$, and by $d_{n}$ the associated multiplicity, one has the general expression for the thermal part
of the free energy
\be
\bar F(\beta) = \sum_{n}d_{n}\log(1-e^{-\beta E_{n}})\,.
\ee
and the zeta-function regularized value of the Casimir energy
\be
\la{4.7}
E_{c} = \frac{1}{2}\sum_{n}d_{n}E_{n} = \frac{1}{2}\zeta_{E}(-1), \qquad \zeta_{E}(s) = \sum_{n}d_{n}E_{n}^{-s}.
\ee
The associated one-loop partition function $Z_{1}$ in (\ref{4.3}) is then 
\be
\la{4.8}
Z_{1} = e^{-\beta E_{c}}\, \Z(q)\,,
\ee
with 
\be
\Z(q) = e^{-\bar F} = \exp\bigg(\sum_{k}d_{k}\sum_{n=1}^{\infty}\frac{1}{n}e^{-n\beta E_{k}}\bigg) = \PE[\Zsp(q)]\,,
\ee
where the single particle partition function is 
\be
\Zsp(q) =  \sum_{n}d_{n}\, q^{E_{n}}\,.
\ee
Notice  that the zeta-function $\zeta_{E}(s)$ in (\ref{4.7}) admits the following expression in terms of $\Zsp(q)$
\be
\la{4.11}
\zeta_{E}(s) = \frac{1}{\Gamma(s)}\int_{0}^{\infty}d\beta\, \beta^{s-1}\, \Zsp(e^{-\beta}).
\ee
In the present context, the partition function $Z_{1}$ is computed on a background with a twist adapted to the index and 
$E_{c}$ is the so-called supersymmetric Casimir energy \cite{Assel:2015nca,Assel:2014tba,Cassani:2014zwa,Bobev:2015kza} (of fluctuations on the D3 brane). The expected precise relation between $Z_{1}$ and the index is 
\be
Z_{1} = e^{-\beta E_{c}}\,\I^{\rm D3}(q)\, .
\ee
Comparing with (\ref{4.8}) gives the key identification
\be
\Z(q) = \I^{\rm D3}(q),
\ee
that clearly holds also at the level of single particle partition function and index. Thus, in our case, we want to check 
\be
\la{4.14}
\sum_{\rm fluctuations}\Zsp(q) = \Isp_{\sX}^{\rm D3}(q) = \frac{1}{q}+1-q\,,
\ee
where we used  (\ref{4.1}). \footnote{We remark that in the unflavored case the three wrapped configurations on $\sX=0$, $\sY=0$, $\sZ=0$ are completely equivalent.}

\subsection{Computation of the single particle partition functions}

\subsubsection{Scalar fluctuations}
\la{sec:scalar-sector}

Let us begin by studying the bosonic scalar fluctuations by expanding at quadratic order the total bosonic action $S_{\rm DBI}+S_{\rm WZ}$.

\paragraph{\underline{$AdS_{5}$ sector}}

It is convenient to introduce new variables $X_{I}$, $I=1, \dots, 4$ in terms of $x, \psi_{i}$
\begin{alignat}{2}
& X_{1} = \sinh x\, \cos\psi_{1}, && X_{2} = \sinh x\, \sin\psi_{1}\, \cos\psi_{2}, \\
& X_{3} = \sinh x\, \sin\psi_{1}\, \sin\psi_{2}\, \cos\psi_{3}, \ \ \ \ &&  X_{4} = \sinh x\, \sin\psi_{1}\, \sin\psi_{2}\, \sin\psi_{3}.
\end{alignat}
One can check that 
\be
dx^{2}+\sinh^{2}x\,(d\psi_{1}^{2}+\sin^{2}\psi_{1}\,d\psi_{2}^{2}+\sin^{2}\psi_{1}\sin^{2}\psi_{2}\,d\psi_{3}^{2}) = \underbrace{dX_{I}^{2}}_{\rm flat}-\underbrace{\sinh^{2}x\, dx^{2}}_{\rm quartic}\,,
\ee
and thus, at quadratic order, 
\ba
ds^{2}_{\widetilde{AdS}_{5}} &= (1+X_{I}^{2}-\kappa^{2}(X_{3}^{2}+X_{4}^{2}))\,d\tau^{2}+dX_{I}^{2}+2i\kappa\,(X_{3}dX_{4}-X_{4}dX_{3})d\tau\, .
\ea
Denoting by $a=2,3,4$ the indices of the $S^{3}$ brane coordinates, we have 
\ba
ds^{2}_{\widetilde{AdS}_{5}} &= [1+X_{1}^{2}+X_{2}^{2}+(1-\kappa^{2})(X_{3}^{2}+X_{4}^{2})+\dot X_{I}^{2}
+2i\kappa\,(X_{3}\dot X_{4}-X_{4}\dot X_{3})
]\,(d\xi^{1})^{2}
+\partial_{a}X_{I}\partial_{b}X_{I}d\xi^{a}d\xi^{b}\lp
+\text{off diagonal terms}\,.
\ea
The DBI action, at quadratic order in the fluctuations, reads
\ba
S_{\rm DBI} = \T_{3}\int d^{4}\xi \sqrt{G}, \qquad ds^{2} = G_{ij}d\xi^{i}d\xi^{j}\,,
\ea
and using 
\be
\delta\sqrt{G^{(0)}+G^{(1)}} = \sqrt{G^{(0)}}+\frac{1}{2}\sqrt{G^{(0)}}\, (G^{(0)})^{ij}G^{(1)}_{ij}+\cdots, 
\ee
we get (denoting by $d^{3}\xi$ the $S^{3}$ coordinates),
\be
\delta S_{\rm DBI} = \frac{1}{2}\beta\T_{3}\int d^{3}\xi\sqrt{g}[g^{ab}
\partial_{a}X_{I}\partial_{b}X_{I}+
X_{1}^{2}+X_{2}^{2}+(1-\kappa^{2})(X_{3}^{2}+X_{4}^{2})+\dot X_{I}^{2}
+2i\kappa\,(X_{3}\dot X_{4}-X_{4}\dot X_{3})]\,.
\ee
We now  rescale fluctations by $(\beta\T_{3})^{1/2}$, and set $\partial_{\tau}\to i\frac{2\pi}{\beta}n$, with $n\in \mathbb{Z}$, \ie expand in 
periodic modes on the thermal circle.
The fields $X_{1},X_{2}$ are scalar fields on $S^{3}$ with common squared mass
\be
X_{1}, X_{2}: \quad M^{2} = 1+n_{\beta}^{2}, \qquad n_{\beta} = \frac{2\pi n}{\beta}, \qquad n = 0, \pm 1, \pm 2, \dots.
\ee
Notice that   the squared mass of  a conformally coupled scalar on $S^{1}\times S^{d}$ is $\frac{1}{4}(d-1)^{2}$ that is 1 for $S^{3}$. This is precisely what we have for $n=0$.

The fields $X_{3},X_{4}$ can be written in complex basis $\phi=\frac{X_{3}+iX_{4}}{\sqrt 2}$, $\bar\phi=\frac{X_{3}-iX_{4}}{\sqrt 2}$ and the expansion of the DBI action gives for them 
\be
\delta S_{\rm DBI}^{(X_{3},X_{4})} = \beta\T_{3}\int d^{3}\xi\sqrt{g}[g^{ab}\partial_{a}\bar\phi\partial_{b}\phi
+\bar\phi((1-\kappa^{2})\phi+2\kappa\dot\phi-\ddot\phi)]\,.
\ee
This is a complex scalar on $S^{3}$ with mass
\be
X_{3}, X_{4}: \quad M^{2} = 1+(n_{\beta}+i\kappa)^{2}\,.
\ee
In this case, the conformally coupled scalars have  an imaginary shift in $n_{\beta}$ in the Kaluza-Klein mass term, similar to what happens in the M5 brane case studied in \cite{Beccaria:2023cuo}.

\paragraph{$S^{5}$ sector}
\la{sec:s5-sector}

In the sphere sector we parametrize fluctuations by setting  
\be
w = \arccos\frac{1-\frac{1}{4}(Y_{1}^{2}+Y_{2}^{2})}{1+\frac{1}{4}(Y_{1}^{2}+Y_{2}^{2})}, \qquad z = \arctan\frac{Y_{2}}{Y_{1}}\,,
\ee
so that 
\be
dw^{2}+\sin^{2}w\, dz^{2} = \frac{dY_{1}^{2}+dY_{2}^{2}}{[1+\frac{1}{4}(Y_{1}^{2}+Y_{2}^{2})]^{2}}\,,
\ee
and
\ba
ds^{2}_{S^{5}} &=[ (1-Y_{c}Y_{c})g_{ab}+\partial_{a}Y_{c}\partial_{b}Y_{c}]\,d\xi^{a}d\xi^{b}
+(\dot Y_{c}\dot Y_{c})(d\xi^{1})^{2}+\text{off diagonal terms}\,.
\ea
This gives the following action from DBI at quadratic order 
\be
\delta S_{\rm DBI} = \frac{1}{2}\beta\T_{3}\int d^{3}\xi \sqrt{g}\,[g^{ab}\partial_{a}Y_{c}\partial_{b}Y_{c}
+\dot Y_{c}\dot Y_{c}-3Y_{c}Y_{c}]\,.
\ee
The  Wess-Zumino term is 
\ba
S_{\rm WZ} &= -i\T_{3}\int \cos^{4}w \,dz\wedge \text{vol}_{S^{3}} 
= -i\T_{3}\int  [1-2(Y_{c}Y_{c})]\frac{Y_{1}dY_{2}-Y_{2}dY_{1}}{Y_{c}Y_{c}}\wedge \text{vol}_{S^{3}}\,,
\ea
and thus the contribution to the fluctuations action is 
\ba
\delta S_{\rm WZ} = \beta\T_{3}\int d^{3}\xi\sqrt{g}(4i\, Y_{1}\dot Y_{2})\,.
\ea
The total $\delta S = \delta S_{\rm DBI}+\delta S_{\rm WZ}$ reads
\ba
\delta S = \frac{1}{2}\beta\T_{3}\int d^{3}\xi\sqrt{g}[
g^{ab}\partial_{a}Y_{c}\partial_{b}Y_{c}
+\dot Y_{c}\dot Y_{c}
-3Y_{c}Y_{c}+8i\,Y_{1}\dot Y_{2}
]\,.
\ea
Introducing again complex combinations $\eta=\frac{Y_{1}+iY_{2}}{\sqrt 2}$, $\bar\eta=\frac{Y_{1}-iY_{2}}{\sqrt 2}$, we get 
\ba
\delta S = \beta\T_{3}\int d^{3}\xi\sqrt{g}[
g^{ab}\partial_{a}\bar\eta\partial_{b}\eta+\bar\eta(-3\eta+4\dot\eta-\ddot\eta)]\,.
\ea
This corresponds to a complex field $\eta$ with the following squared mass 
\be
M^{2} = 1+(n_{\beta}+2i)^{2}\,.
\ee

\subsubsection{Fermionic fluctuations}

At quadratic order in the fermions, the Euclidean fermionic action of the D3 brane is decoupled from the gauge fields and reads \cite{Marolf:2003ye,Marolf:2003vf,Marolf:2004jb}
\be
S_{F} = \frac{1}{2}\T_{3}\int d^{4}\xi\, \sqrt{\det g}\, \overline\Theta(1-\Gamma_{\rm D3})\Gamma^{\alpha}D_{\alpha}\Theta\,,
\ee
where $\Theta$ is a doublet of 10d positive chirality Majorana-Weyl spinors, $\Gamma_{\alpha}=\partial_{\alpha} x^{m}\Gamma_{m}$, $D_{\alpha}=\partial_{\alpha} x^{m}D_{m}$
are pullbacks, and $\Gamma_{D3}$ is a projector required by $\kappa$-symmetry invariance. The 10d covariant derivative is 
\be
D_{m}=\nabla_{m}+\frac{1}{16}\slashed{F}_{(5)}\Gamma_{m}\otimes (i\sigma_{2})\,,
\ee
with $i\sigma_{2}$ acting in doublet space. Let us use Latin letters for 10d indices $0,1,\dots, 9$, Greek letters for D3 brane worldvolume indices $1,2,3,4$, and underlined  indices for flat tangent directions $\hat{0}, \hat{1}, \dots$. In particular \footnote{
$\Gamma_{\alpha} = \partial_{\alpha}X^{m}E\indices{^{\ul{m}}_{m}}\Gamma_{\ul{m}}$, where $E\indices{^{\ul{m}}_{n}}$ is the vielbein.}
\be
\Gamma_{\rm D3} = \frac{\eps^{\alpha_{1}\dots\alpha_{4}}}{4!\sqrt{\det g}}\Gamma_{\alpha_{1}\dots\alpha_{4}} = \Gamma_{\ulh{0}\ulh{1}\ulh{2}\ulh{3}\ulh{4}}\,.
\ee
The two spinors in the doublet $\Theta = (\theta_{1}, \theta_{2})$ obey $\Gamma_{10}\theta_{i}=\theta_{i}$ with $\Gamma_{10} = \Gamma_{\ul{\hat{0}\dots\hat{9}}}$. We fix $\kappa$-symmetry by 
$(1-\Gamma_{10}\otimes \sigma_{3})\Theta=0$ or $\theta_{2}=0$ and we will denote $\theta\equiv \theta_{1}$. Using $\{\Gamma_{\rm D3},\Gamma_{\alpha}\}=0$, we get 
\be
S_{F} = \frac{1}{2}\T_{3}\int d^{4}\xi\, \sqrt{\det g}\, \bar\theta\, \Gamma^{\alpha}\,\bigg(D_{\alpha}+\frac{1}{16}\Gamma_{\rm D3}\slashed{F}_{5}\Gamma_{\alpha}\bigg)\theta\,.
\ee
Our coordinates are 
\bea
ds^{2}_{\widetilde{AdS}_{5}} &= dx^{2}+\sinh^{2}x\, d\widetilde{S}_{3}+\cosh^{2}x\, d\tau^{2}\,,  \qquad \tau=\tau+\beta\,, \\
d\widetilde{S}_{3} &= d\psi_{1}^{2}+\sin^{2}\psi_{1}\,d\psi_{2}^{2}+\sin^{2}\psi_{1}\sin^{2}\psi_{2}\,(d\psi_{3}+i\kappa d\tau)^{2}\,,\\
ds^{2}_{ S^{5}} &= dw^{2}+\cos^{2}w\, dS_{3}+\sin^{2}w\,dz^{2}\,.
\eea
with $\xi^{\alpha}$ being $\tau$ and the coordinates of $S^{3}$. We label them as 
\be
\def\arraystretch{1.3}
\begin{array}{c|ccc|ccc}
X^{m} & x & \psi_{1}, \psi_{2}, \psi_{3} & \tau = \xi_{1} & w & S_{3} = (\xi^{2},\xi^{3},\xi^{4}) & z \\
\midrule
m & 0 & 6,7,8  & 4 & 5 & 1,2,3 & 9
\end{array}\notag
\ee
On the classical solution, that has in particular $x=0$, we remark the non-zero spin-connection
$\Omega^{\ul{\hat{4}\hat{0}}}_{4} = \cosh x$ that  will play an important role. 
The induced covariant derivative is 
\be
D_{\alpha} = \partial_{\alpha}X^{m}D_{m} = \partial_{\alpha}X^{m}\bigg(\partial_{m}+\frac{1}{4}\Omega^{\ul{ab}}_{m}\Gamma_{\ul{ab}}\bigg)\,.
\ee
For $m=1,2,3,4$ we have $\partial_{\alpha}X^{m}=\delta^{m}_{\alpha}$ on the classical solution, so 
\be
D_{\alpha} = \partial_{\alpha}+\frac{1}{4}\Omega^{\ul{ab}}_{\alpha}\Gamma_{\ul{ab}}\,.
\ee
For $\alpha=1,2,3$ on classical solution we obtain just the covariant derivative on $S^{3}$. For $\alpha=4$
\be
D_{4} = \partial_{4}+\frac{1}{2}\Gamma_{\ulh{4}\ulh{0}}\,.
\ee
Hence, 
\be
S_{F} = \frac{1}{2}\T_{3}\int d^{4}\xi\, \sqrt{\det g}\, \bar\theta\, (\slashed{D}+M)\theta\,.
\ee
where
\ba
M &= \Gamma^{\ulh{4}}\frac{1}{2}\Gamma_{\ulh{4}\ulh{0}}\Gamma_{\ulh{4}}-\frac{1}{16}\Gamma_{\rm D3}\Gamma^{\alpha}\slashed{F}_{5}\Gamma_{\alpha}
=  \frac{1}{2}\Gamma_{\ulh{0}\ulh{4}}-\frac{1}{16}\Gamma_{\rm D3}\Gamma^{\alpha}\slashed{F}_{5}\Gamma_{\alpha}\,.
\ea
Around the classical solution $\Gamma_{\alpha}=\Gamma^{\alpha}$. Using the 5-form background expression, we get  
\be
\slashed{F}_{5} = -4(\Gamma^{\ulh{0}\ulh{6}\ulh{7}\ulh{8}\ulh{4}}+\Gamma^{\ulh{5}\ulh{1}\ulh{2}\ulh{3}\ulh{9}})\,.
\ee
Thus,
\bea
\sum_{\alpha=1}^{3}\Gamma^{\alpha}\slashed{F}_{5}\Gamma_{\alpha} &=  -4\times 3
(-\Gamma^{\ulh{0}\ulh{6}\ulh{7}\ulh{8}\ulh{4}}+\Gamma^{\ulh{5}\ulh{1}\ulh{2}\ulh{3}\ulh{9}})\,, \qquad
\Gamma^{\ulh{4}}\slashed{F}_{5}\Gamma_{\ulh{4}} =  -4
(\Gamma^{\ulh{0}\ulh{6}\ulh{7}\ulh{8}\ulh{4}}-\Gamma^{\ulh{5}\ulh{1}\ulh{2}\ulh{3}\ulh{9}})\,.
\eea
The chirality constraint gives
\be
\Gamma_{\ulh{0}\ulh{1}\ulh{2}\ulh{3}\ulh{4}\ulh{5}\ulh{6}\ulh{7}\ulh{8}\ulh{9}}\theta=\theta, \quad 
\Gamma_{\ulh{0}\ulh{6}\ulh{7}\ulh{8}\ulh{4}\ulh{5}\ulh{1}\ulh{2}\ulh{3}\ulh{9}}\theta=-\theta, \quad 
\Gamma_{\ulh{5}\ulh{1}\ulh{2}\ulh{3}\ulh{9}}\theta=-\Gamma_{\ulh{4}\ulh{8}\ulh{7}\ulh{6}\ulh{0}}\theta = \Gamma_{\ulh{0}\ulh{6}\ulh{7}\ulh{8}\ulh{4}}\theta\,,
\ee
and $
\Gamma^{\alpha}\slashed{F}_{5}\Gamma_{\alpha} = 16 \Gamma^{\ulh{0}\ulh{6}\ulh{7}\ulh{8}\ulh{4}}$.
The mass term is therefore
\be
M = \frac{1}{2}\Gamma_{\ulh{0}\ulh{4}}-\Gamma_{\ulh{0}\ulh{1}\ulh{2}\ulh{3}\ulh{4}}\Gamma^{\ulh{0}\ulh{6}\ulh{7}\ulh{8}\ulh{4}} = 
\frac{1}{2}\Gamma_{\ulh{0}\ulh{4}}+\Gamma_{\ulh{1}\ulh{2}\ulh{3}\ulh{6}\ulh{7}\ulh{8}} = \frac{1}{2}\Pi_{1}+\Pi_{2}\,,
\ee
with commuting $\Pi_{1,2}$ with $\Pi_{1,2}^{2}=-\mb{I}$. 

In conclusion, expanding fermions in Fourier modes along the thermal cycle, $\partial_{1}\to in_{\beta}$, and diagonalizing the two projectors, 
we obtain the Dirac operators
\be
\mc D_{F} = i\Gamma^{i}\nabla_{i}+M, \qquad M = n_{\beta}+i\nu, \qquad \nu = \pm\frac{1}{2} , \pm \frac{3}{2}\,.
\ee
If needed, the squared fermionic operator is 
\be
\Delta_{\frac{1}{2}} = -\nabla^{2}_{S^{3}}+\frac{1}{4}R+M^{2} = -\nabla^{2}_{S^{3}}+\frac{3}{2}+M^{2}\,,
\ee
where $\nabla_{S^{3}}$ contains the spinor connection and we used the scalar curvature $R(S^{1}\times S^{3}) = R(S^{3}) = 6$.

\subsubsection{Gauge field fluctuations}

At quadratic order, the gauge field decouples from the other fluctuations and its action is the Maxwell one $S=-\frac{1}{4}\int d^{4}\xi \sqrt{g}F^{2}$
on $S^{1}_{\beta}\times S^{3}$. At this order, the background twist is not seen.  So, we don't need 
any shift in its mass after expansion in $S^{1}_{\beta}$ modes.
As expected from supersymmetry, the field strength $F_{ab}$ together with the 6 scalars and 4 Majorana-Weyl fermions
is part of an Abelian  $\N=4$ Maxwell multiplet. 

\subsection{Unflavored Schur index from twisted $\N=4$ fields on $S^{1}_{\beta}\times S^{3}$}
\la{sec:index-twist}

For the fields in the  $\N=4$ Maxwell multiplet on $S^{1}_{\beta}\times S^{3}$, including values of the shifts in the mass term for scalars and fermions we found 
the values summarized in Table~\ref{tab:1}.
%\begin{table}[H]
%\be
%\def\arraystretch{1.3}
%\begin{array}{cc}
%\toprule
%\textsc{fluctuations} & \nu  \\
%\midrule
%X_{1}, X_{2} & 0 \\
%X_{3}, X_{4} & 1 \\
%\midrule
%Y_{1},Y_{2} & 2 \\
%\midrule
%\psi & \frac{1}{2}, \frac{1}{2}, \frac{3}{2}, \frac{3}{2} \\
%\midrule
%V_{a} & 0\\
%\bottomrule
%\end{array}\notag
%\ee
%\caption{Shifts for the $S^{1}_{\beta}$ reduced mass term of conformally coupled fluctuations on $S^{1}_{\beta}\times S^{3}$.
%\la{tab:1}
%}
%\end{table}
\begin{table}[H]
\be
\def\arraystretch{1.3}
%\begin{array}{c|cc|c|c|c}
\begin{array}{cccccc}
\toprule
\textsc{fluctuations} & X_{1}, X_{2} & X_{3},X_{4} & Y_{1},Y_{2} & \psi & V_{a}  \\
\midrule
\nu &  0 &  1 &  2 & \frac{1}{2}, \frac{1}{2}, \frac{3}{2}, \frac{3}{2} & 0 \\
\bottomrule
\end{array}\notag
\ee
\caption{Shifts for the $S^{1}_{\beta}$ reduced mass term of conformally coupled fluctuations on $S^{1}_{\beta}\times S^{3}$.
\la{tab:1}
}
\end{table}
The associated single particle partition functions can be obtained for $\nu=0$ by using conformal symmetry (and the absence of conformal anomaly in the $S^{1}\times S^{3}$ case)
and mapping the problem to operator counting on $\mathbb R^{4}$ \cite{Cardy:1991kr,Kutasov:2000td}. The $\nu>0$ case corresponds to a simple dressing by the 
universal factor $\frac{1}{2}(q^{\nu}+q^{-\nu})$, see \cite{Beccaria:2023cuo} for the completely analogous 6d calculation. This gives 
\bea
\la{6.1}
\Zsp^{\phi}(q;\nu) &= \frac{q(1-q^{2})}{(1-q)^{4}}\frac{q^{\nu}+q^{-\nu}}{2}\,,  \\
\Zsp^{\psi}(q;\nu) &= \frac{4q^{3/2}(1-q)}{(1-q)^{4}}\frac{q^{\nu}+q^{-\nu}}{2}\,,  \\
\Zsp^{V}(q) &= \frac{2q^{2}(1-q)(3-q)}{(1-q)^{4}}\,.
\eea
To recall, in (\ref{6.1}) the common denominator $(1-q)^{4}$ corresponds to the application in all possible ways of the four derivatives $\partial_{\mu}$ to the 
fields $\phi, \psi, V$. For scalars, the numerator $q-q^{3}$ correspond to the elementary field $\phi$ with a subtraction of the vanishing equation of motion $\partial^{2}\phi=0$.
For the four Majorana-Weyl fermions, the numerator $q^{3/2}-q^{5/2}$ corresponds similarly to the elementary field $\psi$ with subtraction of $\slashed{\partial}\psi=0$.
For the gauge field, the terms in $2q^{2}(1-q)(3-q) = 6q^{2}-8q^{3}+2q^{4}$ have a similar origin. We have six gauge invariant fields $F_{\mu\nu}$, the subtraction of 
 vanishing $4+4$ terms from  $\partial^{\mu}F_{\mu\nu}=0$ and $\partial^{\mu}F^{\star}_{\mu\nu}=0$, and the double subtraction (thus with positive sign) of the scalar quantities
 $\partial^{\mu}\partial^{\nu}F_{\mu\nu}=\partial^{\mu}\partial^{\nu}F^{\star}_{\mu\nu}=0$.

Using data in Table~\ref{tab:1}, we get (including the fermion sign)
%\ba
%\la{6.2}
%\Zsp(q) &= 2[\Zsp^{\phi}(q;0)+\Zsp^{\phi}(q;1)+\Zsp^{\phi}(q;2)]-2[\Zsp^{\psi}(q;\tfrac{1}{2})+\Zsp^{\psi}(q;\tfrac{3}{2})]+\Zsp^{V}(q) \lp
%= 1+\frac{1}{q}-q \,.
%\ea
\ba
\la{6.2}
 \Zsp(q) = \Zsp^{\rm scalars}(q)+\Zsp^{\rm fermions}(q)+\Zsp^{V}(q)  = 1+\frac{1}{q}-q \,, 
\ea
with
\bea
\Zsp^{\rm scalars}(q) &= 2[\Zsp^{\phi}(q;0)+\Zsp^{\phi}(q;1)+\Zsp^{\phi}(q;2)]\,, \\
\Zsp^{\rm fermions}(q) &= -2[\Zsp^{\psi}(q;\tfrac{1}{2})+\Zsp^{\psi}(q;\tfrac{3}{2})]\,.
\eea
The equality in (\ref{6.2}) is in agreement with (\ref{4.14}). The constant term in (\ref{6.2}) leads to a formal divergence in the index after plethystic exponentiation. It originates from an effective total of 
two bosonic zero modes. Recall that each bosonic/fermionic zero mode contributes a constant term $\pm \frac{1}{2}$ to the total $\Z(q)$. From the expansion of expressions (\ref{6.1}), 
one finds the zero mode counting in Table \ref{tab:1bis}.
\begin{table}[H]
\be
\def\arraystretch{1.3}
%\begin{array}{c|cc|c|c|c}
\begin{array}{ccccc|cc|c}
\toprule
\textsc{fluctuations} & X_{1}, X_{2} & X_{3},X_{4} & Y_{1},Y_{2} & V^{a} & \psi^{(1/2)} & \psi^{(3/2)} & \text{total}  \\
\midrule
\#\, \text{zero modes} &  0 &  2 &  8 & 0 & 0 & -8 & 2 \\
\bottomrule
\end{array}\notag
\ee
\caption{Number of zero modes for the various bosonic and fermionic fluctuations. Fermionic fluctiations are labeled with their shift $\psi^{(\nu)}$ and the negative sign is the 
$(-1)^{\rm F}$ to be taken into account in total.
\la{tab:1bis}
}
\end{table}
As in the M5 case discussed in \cite{Beccaria:2023cuo}, with the standard normalization of the D3 brane path integral, each zero mode is expected 
to contribute a factor of $\T_{3}^{1/2}\sim \sqrt{N}$. This explains the peculiar ``wall-crossing'' enhancement factor of $N$ in (\ref{2.15}), from the total of 2 zero modes. 
This factor is fully regularized in the index expression 
with non-trivial fugacity $u$, as discussed in next section.

\subsection{Supersymmetric Casimir energy on the wrapped brane}

Using (\ref{4.11}) for the partition functions in (\ref{6.1}), 
%\footnote{An alternative simpler computation is based on the high temperature expansion
%$-\frac{1}{2}\frac{d}{d\beta}\Zsp(e^{-\beta}) = \text{singular terms}+E_{c}+\mc O(\beta)$ that gives same as using (\ref{4.11}). \red{[this is true but why ? does it hold in general ???]}}
we compute the following supersymmetric Casimir energies \footnote{Strictly speaking, what is relevant here is the sum over all fields 
belonging to the Maxwell multiplet. Still, it may be of interest to give the individual fluctuation contribution to the total.}
\be
E_{c}^{\phi}(\nu) = \frac{1-10\nu^{4}}{240}, \qquad E_{c}^{\psi}(\nu) = -\frac{17-120\nu^{2}+80\nu^{4}}{960}, \qquad E_{c}^{V} = \frac{11}{120},
\ee
that reduce to the standard values for $\nu=0$, see for instance Table 2 in \cite{Beccaria:2014xda} (there, the minus sign for fermions is not incorporated in $E_{c}$).
The total is 
\be
\la{6.5}
E_{c}^{\rm D3} = 2E_{c,\phi}(0)+2E_{c,\phi}(1)+2E_{c,\phi}(2)-2E_{c,\psi}(\tfrac{1}{2})-2E_{c,\psi}(\tfrac{3}{2})+E_{c,V} = -1,
\ee
which is of course also immediately obtained using the total single particle function in (\ref{6.2}). This supersymmetric Casimir energy is equivalently obtained
as, \cf (\ref{2.1}),
\be
\la{6.6}
E_{c}^{\rm D3} = \frac{1}{2}\zeta^{\rm D3}(-1), \qquad \zeta^{\rm D3}(s) = \sum(-1)^{F}(H+J+\bar J)^{-s}\,.
\ee
where the sum is over the single particle states of the modes on the wrapped D3 brane 
and should not be confused with the same quantity for $\N=4$ SYM (hence the ``D3'' label).
In more details, the states are derived in Appendix A of \cite{Imamura:2021ytr} that we summarize  in Table \ref{tab:states} for the  convenience of the reader.
\begin{table}[H]
\be
\def\arraystretch{1.3}
\begin{array}[t]{cccccc}
\toprule
\textsc{state} & H & J & \bar J & R & \bar R \\
\midrule
\bar\phi & \ell-1 & 0 & 0 & \frac{\ell}{2} & \frac{\ell}{2} \\
\bar\chi & \ell-\frac{1}{2} & 0 & \frac{1}{2} & \frac{\ell}{2} & \frac{\ell-1}{2} \\
F^{-} & \ell & 0 & 0 & \frac{\ell}{2} & \frac{\ell-2}{2} \\
\psi & \ell-\frac{1}{2} & \frac{1}{2} & 0 & \frac{\ell-1}{2} & \frac{\ell}{2} \\
w & \ell & \frac{1}{2} & \frac{1}{2} & \frac{\ell-1}{2} & \frac{\ell-1}{2} \\
\bar\psi & \ell+\frac{1}{2} & \frac{1}{2} & 0 & \frac{\ell-1}{2} & \frac{\ell-2}{2} \\
%F^{+} & \ell & 0 & 0 & \frac{\ell-2}{2} & \frac{\ell}{2} \\
%\chi & \ell+\frac{1}{2} & 0 & \frac{1}{2} & \frac{\ell-2}{2} & \frac{\ell-1}{2} \\
%\phi & \ell+1 & 0 & 0 & \frac{\ell-2}{2} & \frac{\ell-2}{2}\\
\bottomrule
\end{array}\quad
\begin{array}[t]{cccccc}
\toprule
\textsc{state} & H & J & \bar J & R & \bar R \\
\midrule
%\bar\phi & \ell-1 & 0 & 0 & \frac{\ell}{2} & \frac{\ell}{2} \\
%\bar\chi & \ell-\frac{1}{2} & 0 & \frac{1}{2} & \frac{\ell}{2} & \frac{\ell-1}{2} \\
%F^{-} & \ell & 0 & 0 & \frac{\ell}{2} & \frac{\ell-2}{2} \\
%\psi & \ell-\frac{1}{2} & \frac{1}{2} & 0 & \frac{\ell-1}{2} & \frac{\ell}{2} \\
%w & \ell & \frac{1}{2} & \frac{1}{2} & \frac{\ell-1}{2} & \frac{\ell-1}{2} \\
%\bar\psi & \ell+\frac{1}{2} & \frac{1}{2} & 0 & \frac{\ell-1}{2} & \frac{\ell-2}{2} \\
F^{+} & \ell & 0 & 0 & \frac{\ell-2}{2} & \frac{\ell}{2} \\
\chi & \ell+\frac{1}{2} & 0 & \frac{1}{2} & \frac{\ell-2}{2} & \frac{\ell-1}{2} \\
\phi & \ell+1 & 0 & 0 & \frac{\ell-2}{2} & \frac{\ell-2}{2}\\
\bottomrule
\end{array}\notag
\ee
\caption{Single particle states on the wrapped D3 brane and their quantum numbers.
A value of $J$ corresponds of course to a full $SU(2)$ multiplet with $2J+1$ values of $J_{z}$, and the same for $\bar J, R, \bar R$.
\la{tab:states}
}
\end{table}
These fields are the modes on the $\sZ=0$ brane. They are obtained decomposing the Maxwell multiplet with respect to the unbroken supersymmetry on the  brane wrapped on $S^{1}\times S^{3}$.
In particular, they are organized in multiplet of the bosonic algebra $SU(2)_{J}\times SU(2)_{\bar J}\times SU(2)_{R}\times SU(2)_{\bar R}$ (we omit $U(1)$ factors not relevant for our discussion).
The index $\ell=0, 1, 2, \dots$  represents the angular momentum after Kaluza-Klein expansion in $S^{3}$ spherical harmonics. 
Summing over the states in the table we get 
\be
\zeta^{\rm D3}(s) = S_{\ell=0}(s) +S_{\ell=1}(s) +\sum_{\ell=2}^{\infty}S_{\ell}(s), \qquad S_{\ell}(s) = \sum_{{\rm fixed\ }\ell} (-1)^{F}(H+J+\bar J)^{-s}\,,
\ee
where $S_{\ell}(s)$ is a finite sum and we need to split $\ell=0$ and $\ell=1$ because representations with negative $R$ or $\bar R$ should be omitted. From Table \ref{tab:states} we compute
\be
\la{6.8}
\zeta^{\rm D3}(-1) = -1+0+\lim_{s\to -1}\sum_{\ell=2}^{\infty}[(\ell-1)^{-s}-2\ell^{-s}+(\ell+1)^{-s}] = -1+\lim_{s\to -1}(1-2^{-s}) = -2.
\ee
Plugging this in (\ref{6.6}) we match (\ref{6.5}). Of course this had to work a priori, but the aim of the above discussion is to emphasize the origin of the supersymmetric Casimir energy
as a regularized sum of the combination of $H$ and $J, \bar J$ appearing in the index, see also \cite{Kim:2012ava,Kim:2012qf,Kim:2013nva} for wrapped M5 brane in the context of the 6d $(2,0)$ theory.
Notice that the peculiar $1/q$ term in (\ref{6.2}), or ``$-1$'' term in (\ref{6.8}), comes from the $\ell=0$ mode of the scalar $\bar \phi$. This negative energy state is not a tachyon and is compatible with unbroked supersymmetry
due to the non-trivial geometry of the wrapped brane. 

\section{Flavored Schur index}
\la{sec:flavor}

In our approach, the introduction of flavor fugacities in the superconformal index does not pose conceptual problems. Still, it is instructive to 
work out the details and, in particular, how the extra fugacities are included in the operator counting partition functions (\ref{6.1}). 
Consistently with the constraint in  (\ref{2.1}), we take 
\be
(u_{1},u_{2},u_{3})=(u, u^{-1}, 1)\,. 
\ee
As a first remark, the effect of the twist related to the $J, \bar J$ charges in (\ref{2.5}) does not interfere with the 
$u$-dependent twists in the $S^{3}$ part of the brane worldvolume for the configuration $\sZ=0$.
Indeed, as discussed in \cite{Beccaria:2023cuo} in the case of the M5 brane system, 
in a static gauge the $\nu$ shift does not depend on $u$ that can be absorbed in the non-fluctuating coordinates of $S^{3}$. This will not be true for the $\sX=0$ and $\sY=0$ configurations
and will lead to a $u$ dependent correction to the shift $\nu$.

Apart from the set of shifts $\nu$, what remains to be found is the  suitable $u$-dependent generalization of the single particle partition functions in (\ref{6.1}) obtained in principle from the 
study of the $\N=4$ Maxwell multiplet fields 
on the 4d space $S^{1}_{\beta}\times \widetilde{S}^{3}$ where $\widetilde{S}^{3}$ is obtained from  $S^{5}$ with twisted metric
\be
ds^{2}_{\widetilde{S}^{5}} = \sum_{i=1}^{3}\bigg[dn_{i}^{2}+n_{i}^{2}(d\varphi_{i}+i\alpha_{i}\,d\tau)^{2}\bigg], \qquad \sum_{i=1}^{3}n_{i}^{2}=1\,,
\ee
where the three mixing coefficients $\alpha_{i}$ are related to $u_{i}$  by $q^{\alpha_{i}} = u_{i}$ and the wrapped configurations  $\sX=0$, $\sY=0$, $\sZ=0$ correspond
to $n_{1}=0$, $n_{2}=0$, $n_{3}=0$, respectively.

A second remark concerns the classical factors $u^{N}$, $u^{-N}$ in (\ref{3.14}). Given a wrapping with $n_{i}=0$, the 4-form (\ref{5.3}) requires $dz\to dz+i\alpha \,d\tau$ where $z$ is identified with the angle 
$\varphi_{i}$ and 
$\alpha \equiv \alpha_{i}$. Evaluation of the classical D3 action reproduces those factors from the Wess-Zumino part $S^{\rm cl}_{\rm WZ}$.

\subsection{Single particle partition functions of twisted fields in $S^{1}_{\beta}\times S^{3}$}

What remains to be done is to compute the modified $\Zsp(q;\nu)\to \Zsp(q,u;\nu)$ in (\ref{6.1}). Again, this has a simple operator counting derivation. Before discussing that, let us work out explicitly the result for 
a scalar field to show the structure of the expected result. By a relabeling of coordinates we can write the twisted $\widetilde{S}^{3}$ metric (in all configurations) as 
\be
ds^{2}_{\widetilde{S}^{3}} =dn_{1}^{2}+dn_{2}^{2}+n_{1}^{2}(d\varphi_{1}+i\alpha d\tau)^{2}+n_{2}^{2}(d\varphi_{2}+i\alpha' d\tau)^{2}, \qquad n_{2}^{1}+n_{2}^{2}=1\,.
\ee
Changing coordinates, this gives the metric of $S^{1}_{\beta}\times \widetilde{S}^{3}$  as 
\be
ds^{2}_{^{1}_{\beta}\times \widetilde{S}^{3}} = d\tau^{2}+d\chi^2 +\cos^2 \chi\, (d\varphi_{1}+i\alpha d\tau)^{2}
+\sin^2 \chi\, (d\varphi_{2}+i\alpha' d\tau)^2\,.
\ee
A convenient basis for the eigenfunctions  of the Laplacian on the standard  $S^3$ is  (see, e.g., \cite{lehoucq2003eigenmodes,lachieze2004laplacian})
%%%%%%%%%%%%%%%%
\be
\la{7.5}
\Phi_{p, r_{1},r_{2}} = f_{p, r_{1}, r_{2}}(\chi)\ e^{i(r_{1}+r_{2})\varphi_{1}}\ e^{i(r_{2}-r_{1})\varphi_{2}},\qquad  
r_{1},r_{2}=-\ha {p}, \dots, \ha{p} \ , \ \ \   p=0,1, ...  \ . 
\ee
Using this representation for the wave-functions after a redefinition of $\varphi_{1}, \varphi_{2}$ that momentarily absorbs $\tau$ in the mixing terms, we get 
the scalar free energy 
 \footnote{For the standard Laplacian on $S^3$ 
 we have  $\lambda_p = p(p+2)$    and ${\rm d}_p= (p + 1)^2 $ 
 and for a conformally coupled scalar  we need to add 1 to $\lambda_p $.}
 \bea
F &= \tfrac{1}{2}\log\det(-\nabla^{2}  +1)  \\
& = \tfrac{1}{2}\sum_{p=0}^{\infty}\, \sum_{r_{1},r_{2}=-p/2}^{p/2}
\log \Big[\big(\nb+i\alpha (r_{1}+r_{2})+i\alpha'(r_{1}-r_{2})\big)^{2}+p(p+2)+1\Big]\,.
\eea
The  corresponding  thermal partition function  and the single-particle partition function  are then 
\be
\bar F =  \sum_{p=0}^{\infty}\, \sum_{r_{1},r_{2}=-p/2}^{p/2}
\log[1-e^{-\beta(p+1+ (\alpha(r_{1}+r_{2})+\alpha'(r_{1}-r_{2}))}]  
\ee
and
\be
\la{7.8}
\Zsp^{\phi}(q,u,u') =  \sum_{p=0}^{\infty}\, \sum_{r_{1},r_{2}=-p/2}^{p/2}q^{p+1+\alpha(r_{1}+r_{2})+\alpha'(r_{1}-r_{2})} = 
\frac{q-q^{3}}{(1-u\,q)(1-u^{-1}\,q)(1-u'\,q)(1-u'^{-1}\,q)}\,,
\ee
where we denoted
\be
u = q^{\alpha}, \qquad u' = q^{\alpha'}.
\ee
This expression admits a simple derivation from operator counting in  $\mathbb R\times \widetilde{S}^{3}$ where we assign the weights $u, u^{-1}$  to two vector indices
and $u', u'^{-1}$ to the other two. For a scalar field, the numerator of (\ref{7.8}) corresponds to $\phi$ and the subtraction of the equations of motion $\partial^{2}\phi$.
Both are scalar and no factors $u,u'$ are necessary. The denominator factors corresponds to derivatives $\partial_{\mu}$ and they have one vector index with the above
weights. For a spinor, one finds the same effect in derivatives plus the $u,u'$ factors associated with the spinor indices (same for $\psi$ and $\slashed{\partial}\psi$)
\be
\la{7.10}
\Zsp^{\psi}(q,u, u') = (u^{1/2}+u^{-1/2}) (u'^{1/2}+u'^{-1/2})\,\frac{q^{3/2}(1-q)}{(1-u\,q)(1-u^{-1}\,q)(1-u'\,q)(1-u'^{-1}\,q)}.
\ee
Finally, for the gauge field, the result is 
\bea
\la{7.11}
\Zsp^{V}(q,u, u') &= \frac{c_{1}(u,u')q^{2}-c_{2}(u,u')q^{3}+2q^{4}}{(1-u\,q)(1-u^{-1}\,q)(1-u'\,q)(1-u'^{-1}\,q)}\,, \\
c_{1}(u,u') & = 2+(u+u^{-1})(u'+u'^{-1})\, \\
c_{2}(u,u') &= 2\,(u+u^{-1}+u'+u'^{-1})\,.
\eea
Here, the $q^{2}$ term in the numerator corresponds to six $F_{\mu\nu}$ fields with the $u,u'$ weights carried by the two vector indices
\be
\def\arraystretch{1.3}
\begin{array}{ccccccc}
\text{field} & F_{12} & F_{13} & F_{14} & F_{23} & F_{24} & F_{34} \\
\text{weight} & 1 & uu' & uu'^{-1} & u^{-1}u' & u^{-1}u'^{-1} & 1
\end{array}\notag
\ee
with total equal to $c_{1}(u,u')$. The $q^{3}$ term is the subtraction of overcounted derivatives on the vanishing
vanishing $\partial^{\mu}F_{\mu\nu}=0$ and $\partial^{\mu}F^{\star}_{\mu\nu}=0$. These have one vector index and give total $c_{2}(u,u')$.
Finally, the $2q^{4}$ term accounts for the overcounted derivatives $\partial^{\mu}\partial^{\nu}F_{\mu\nu}=\partial^{\mu}\partial^{\nu}F^{\star}_{\mu\nu}=0$.
These have no free indices and thus no $u,u,'$ weights.

Using the above considerations we can discuss now separately the contribution from the $\sZ=0$ brane, symmetric under $u\to u^{-1}$, and those of the  $\sX=0$, $\sY=0$ branes,  related by  $u\to u^{-1}$, \cf (\ref{3.12}).

\subsection{$\sZ=0$ brane contribution}

Let us consider the D3 brane wrapped on $\sZ=0$. Its single particle index reads \footnote{Use the general expression (\ref{3.4}) after permutation of fugacities or
take $y=v=\eta=1$ in the last equation in Eq.~(4.35) of \cite{Arai:2019xmp}.}
\be
\la{7.12}
\Isp^{\rm D3}_{\sZ}(q,u) = \frac{q^{-1}-1-(u+u^{-1})\,q+3q^{2}-q^{3}}{(1-uq)(1-u^{-1}q)}\,.
\ee
This configuration does not contribute the Schur index,  consistently with \cf (\ref{3.14}). 
This is due to the fermionic term $-1$ in the expansion 
of $\Isp^{{\rm D3}, \sZ=0}$
\be
\Isp^{\rm D3}_{\sZ}(q,u) = q^{-2}+u+u^{-1}-1+\mc O(q)\,,
\ee
that gives a zero after plethystic, see \cite{Arai:2019xmp} for more discussions. Nevertheless, it is instructive to derive (\ref{7.12}) -- thus at the level of single particle index which is non-vanishing.
One can check that the generalization of (\ref{6.2}) is 
\ba
\Zsp^{\rm scalars}_{\sZ}(q,u)+\Zsp^{\rm fermions}_{\sZ}(q,u) +\Zsp^{V}_{\sZ}(q,u) = \Isp_{\sZ}^{\rm D3}(q,u) \,,
\ea
with 
\bea
\Zsp^{\rm scalars}_{\sZ}(q,u) &= 2[\Zsp^{\phi}_{\sZ}(q,u;0)+\Zsp^{\phi}_{\sZ}(q,u;1)+\Zsp^{\phi}_{\sZ}(q,u;2)]\, , \\
\Zsp^{\rm fermions}_{\sZ}(q,u) &= -2[\Zsp^{\psi}_{\sZ}(q,u;\tfrac{1}{2})+\Zsp^{\psi}_{\sZ}(q,u;\tfrac{3}{2})]\, , 
\eea
and 
\bea
\la{7.16}
\Zsp^{\phi}_{\sZ}(q,u;\nu) &= \frac{q(1-q^{2})}{(1-uq)^{2}(1-u^{-1}q)^{2}}\frac{q^{\nu}+q^{-\nu}}{2}\,, \\
\Zsp^{\psi}_{\sZ}(q,u;\nu) &= (u+2+u^{-1})\,\frac{q^{3/2}(1-q)}{(1-uq)^{2}(1-u^{-1}q)^{2}}\frac{q^{\nu}+q^{-\nu}}{2}\,, \\
\Zsp^{V}_{\sZ}(q,u) &= \frac{(u^{2}+4+u^{-2})\,q^{2}-4\,(u+u^{-1})q^{3}+2q^{4}}{(1-uq)^{2}(1-u^{-1}q)^{2}}\,,
\eea
which are the single particle partition functions in (\ref{7.8}), (\ref{7.10}), and (\ref{7.11}), with $u' = u^{-1}$, dressed by the universal factor taking into account the $\nu$ shifts
in Table \ref{tab:1} for  scalars and fermions. \footnote{For the $\sZ=0$ brane, there is no WZ contribution nor DBI mixing $u$-dependent contribution.}

\subsection{$\sX=0$ brane contribution}

In this case, we have to set $u' = 1$ and the single particle partition functions to be used are
\bea
\la{7.17}
\Zsp^{\phi}_{\sX}(q,u; \nu) &= \frac{q(1-q^{2})}{(1-uq)(1-u^{-1}q)(1-q)^{2}}\frac{q^{\nu}+q^{-\nu}}{2}\,, \\
\Zsp^{\psi}_{\sX}(q,u;\nu) &= 2\,(u^{1/2}+u^{-1/2})\,\frac{q^{3/2}(1-q)}{(1-uq)(1-u^{-1}q)(1-q)^{2}}\frac{q^{\nu}+q^{-\nu}}{2}\,, \\
\Zsp^{V}_{\sX}(q,u) &= \frac{2(u+1+u^{-1})\,q^{2}-2\,(u+2+u^{-1})\,q^{3}+2q^{4}}{(1-uq)(1-u^{-1}q)(1-q)^{2}}\,.
\eea 
For this configuration, in the analysis of the  fluctuations in $S^{5}$, we have a non-trivial mixing in the WZ term that will correct the value of $\nu$. This is due to 
$dz \to dz+i\alpha d\tau$ with  $q^{\alpha}=u$  in 
\bea
ds^{2}_{\widetilde{S}^{5}} &= dw^{2}+\cos^{2}w\, dS_{3}+\sin^{2}w\,(dz+i \alpha d\tau)^{2}\,, \\
C_{4} & =  i\T_{3}\,\cos^{4}w\, (dz+i\alpha d\tau)\wedge \text{vol}_{\widetilde{S}^{3}}\,.
\eea
Repeating the calculation in section \ref{sec:scalar-sector} with the replacement $g_{ab}\to \bar{g}_{ab}(u)$ (the twisted $S^{3}$ metric), we have now
\ba
ds^{2}_{\widetilde{S}^{5}} &=[ (1-Y_{c}Y_{c})\bar{g}_{ab}+\partial_{a}Y_{c}\partial_{b}Y_{c}]\,d\xi^{a}d\xi^{b}\lp
+[\dot Y_{c}\dot Y_{c}+2i\alpha (Y_{1}\dot Y_{2}-Y_{2}\dot Y_{1})-\alpha^{2}Y_{c}Y_{c}](d\xi^{1})^{2}+\text{off diagonal terms}\,,
\ea
that gives the following quadratic expansion of the DBI action
\be
\delta S_{\rm DBI} = \frac{1}{2}\beta\T_{3}\int d^{3}\xi \sqrt{\bar g}\,[\bar g^{ab}\partial_{a}Y_{c}\partial_{b}Y_{c}
+\dot Y_{c}\dot Y_{c}+4i\,\alpha\, Y_{1}\dot Y_{2}-(3+\mu^{2})Y_{c}Y_{c}]\,.
\ee
The  WZ term is 
\ba
S_{\rm WZ} &= -i\T_{3}\int \cos^{4}w \,(dz+i\alpha d\tau)\wedge \text{vol}_{\widetilde{S}^{3}} 
= -i\T_{3}\int  (1-2Y_{c}Y_{c})\bigg(\frac{Y_{1}dY_{2}-Y_{2}dY_{1}}{Y_{c}Y_{c}}+i\alpha d\tau\bigg)\wedge \text{vol}_{\widetilde{S}^{3}}\,,
\ea
and thus
\ba
\delta S_{\rm WZ} = \beta\T_{3}\int d^{3}\xi\sqrt{\bar g}\, (4i Y_{1}\dot Y_{2}-2\alpha Y_{c}Y_{c})\,.
\ea
The total $\delta S = \delta S_{\rm DBI}+\delta S_{\rm WZ}$ is 
\ba
\delta S = \frac{1}{2}\beta\T_{3}\int d^{3}\xi\sqrt{\bar g}\,[
\bar g^{ab}\partial_{a}Y_{c}\partial_{b}Y_{c}
+\dot Y_{c}\dot Y_{c}
-(1+\alpha)(3+\alpha)\,Y_{c}Y_{c}+4i(2+\alpha)\,Y_{1}\dot Y_{2}
]\,.
\ea
Introducing $\eta=\frac{Y_{1}+iY_{2}}{\sqrt 2}$, $\bar\eta=\frac{Y_{1}-iY_{2}}{\sqrt 2}$, we get finally
\ba
\delta S = \beta\T_{3}\int d^{3}\xi\sqrt{g}[
g^{ab}\partial_{a}\bar\eta\partial_{b}\eta+\bar\eta(-(1+\mu)(3+\mu)\eta+2(\mu+2)\dot\eta-\ddot\eta)]\, .
\ea
This corresponds to a squared mass for the complex field $\eta$ given by the expressions
\be
M^{2} = 1+(n_{\beta}+i(2+\alpha))^{2}\,.
\ee
A similar $\alpha$ dependent correction to $\nu$ is also present in fermions and the generalization of relation (\ref{6.2}) reads now
%\bea
%& 2[\Zsp^{\phi}_{\sX}(q,u;0)  +\Zsp^{\phi}_{\sX}(q,u;1)+\Zsp^{\phi}_{\sX}(q,u;2+\alpha)] 
% -2[\Zsp^{\psi}_{\sX}(q,u; \tfrac{1}{2}+\alpha)+\Zsp^{\psi}_{\sX}(q,u; \tfrac{3}{2}+\alpha)]
%+\Zsp^{V}_{\sX}(q,u) \\
%&\qquad = \Isp^{\rm D3}_{\sX}(q,u),
%\eea
\ba
\la{7.26}
\Zsp^{\rm scalars}_{\sX}(q,u)+\Zsp^{\rm fermions}_{\sX}(q,u) +\Zsp^{V}_{\sX}(q,u) = \Isp_{\sX}^{\rm D3}(q,u) \,,
\ea
with 
\bea
\Zsp^{\rm scalars}_{\sX}(q,u) &= 2[\Zsp^{\phi}_{\sX}(q,u;0)  +\Zsp^{\phi}_{\sX}(q,u;1)+\Zsp^{\phi}_{\sX}(q,u;2+\alpha)] \, , \\
\Zsp^{\rm fermions}_{\sX}(q,u) &= -2[\Zsp^{\psi}_{\sX}(q,u; \tfrac{1}{2}+\alpha)+\Zsp^{\psi}_{\sX}(q,u; \tfrac{3}{2}+\alpha)]\, , 
\eea
The  r.h.s. in (\ref{7.26}) was given in (\ref{3.10}) and equality is easily checked using the explicit expressions (\ref{7.17}).

 We remark that in this case the small $q$ expansion of the single particle brane index reads
\be
\Isp^{\rm D3}_{\sX}(q,u) = \frac{1}{uq}+\frac{1}{u^{2}}+\mc O(q),
\ee
and the term $1/u^{2}$ admits a finite plethystic (with singular $u\to 1$ limit) regularizing the zero mode contribution discussed in the unflavored case. 

\subsection{Supersymmetric Casimir energy}

It is interesting to look at the supersymmetric Casimir energy on the wrapped brane in presence of the fugacity $u\neq 1$. Let us consider as an illustrative example the 
$\sZ=0$ configuration. We can write the single particle index in (\ref{7.12}) as \footnote{Notice that $q^{1}$ gets two contributions, one from the sum and one from $-q$ before it.}
\be
\Isp_{\sZ}^{\rm D3}(q,u) = \frac{1}{q}+u+u^{-1}-1-q+(1-u)^{2}\,\sum_{p=1}^{\infty}(1+u^{2p})\frac{q^{p}}{u^{p+1}}\,.
\ee
The associated zeta-function on the brane, \cf (\ref{6.6}), is then 
\be
\zeta^{\rm D3}(s) = (-1)^{-s}-1+(1-u)^{2}\,\sum_{p=1}^{\infty}(1+u^{2p})\frac{p^{-s}}{u^{p+1}} = (-1)^{-s}-1+(u+u^{-1}-2)(\text{Li}_{s}(u)+\text{Li}_{s}(u^{-1}))\,.
\ee
Setting $s=-1$ and using $\text{Li}_{-1}(z) = \frac{z}{(1-z)^{2}}$ gives 
\be
\zeta^{\rm D3}(-1) = -2+(u+u^{-1}-2)\bigg(\frac{u}{(1-u)^{2}}+\frac{u^{-1}}{(1-u^{-1})^{2}}\bigg) = 0,\quad \to \quad E_{c}^{\rm D3}=0\,.
\ee
This vanishing implies that the two limits $s\to -1$ and $u\to 1$ do not commute. A similar vanishing is obtained for the $\sX=0$ (and Weyl reflected $\sY=0$) configurations.

\section*{Acknowledgements}

We thank Arkady A. Tseytlin, Yosuke Imamura, and Ji Hoon Lee for useful discussions related to various aspects of this work. 
MB and ACB are supported by the INFN grants GSS and GAST. 

\appendix
\section{Special functions}
\la{app:special}

We collect in this appendix the definition of the special functions appearing in the text and some useful identities.

\paragraph{Dedekind eta function}
\be
\la{A.1}
\eta(\tau) = q^{\frac{1}{12}}\prod_{k=1}^{\infty}(1-q^{2k}), \qquad q=e^{i\pi \tau}\,.
\ee
%Notice that 
%\ba
%\eta(\tau/2) &= q^{1/24}\prod_{k=1}^{\infty}(1-q^{k}) = q^{1/24}\prod_{k=1}^{\infty}(1-q^{2k}) \prod_{k=1}^{\infty}(1-q^{2k-1}) 
%= \eta(\tau)\, q^{-1/12}\, \prod_{k=1}^{\infty}(1-q^{2k-1}).
%\ea

\paragraph{Theta functions}
\bea
\vartheta_{1}(z,q) &= 2q^{1/4}\sin z\, \prod_{k=1}^{\infty}(1-q^{2k})(1-2q^{2k}\cos(2z)+q^{4k})\,, \\
\vartheta_{2}(z,q) &= 2q^{1/4}\cos z\, \prod_{k=1}^{\infty}(1-q^{2k})(1+2q^{2k}\cos(2z)+q^{4k})\,, \\
\vartheta_{3}(z,q) &=  \prod_{k=1}^{\infty}(1-q^{2k})(1+2q^{2k-1}\cos(2z)+q^{4k-2})\,, \\
\vartheta_{4}(z,q) &=  \prod_{k=1}^{\infty}(1-q^{2k})(1-2q^{2k-1}\cos(2z)+q^{4k-2})\,.
\eea

\paragraph{$q$-Pochhammer}

\ba
\la{A.3}
(a,q)_{\infty} = \prod_{k=0}^{\infty}(1-a\,q^{k})\,.
\ea
Notice that we can write the Dedekind function in (\ref{A.1}) as 
\be
\la{A.3}
\eta(\tau) = q^{\frac{1}{12}}(q^{2},q^{2})_{\infty}\,.
\ee

\section{Doubly wrapped D3 brane index in closed form and analytic continuation}
\la{app:w2}

In this appendix we discuss the evaluation of the doubly wrapped D3 brane index $\I_{\sX,2}^{\rm D3}(q,u)$.
We will get it by computing the exact refined index for $\N=4$ $U(2)$ SYM and applying (\ref{3.13}) to the result. This has to be compared with the 
approach in \cite{Arai:2020qaj} where analytic continuation is done at the level of the single particle index, thus requiring some ad hoc procedure to 
evaluate  holonomy integrals.

The refined Schur index in $\mc N=4$ $U(N)$ SYM is given by  the holonomy integral 
\be
\I_{N}^{\rm SYM}(q,u) = \frac{1}{N!}\oint_{|z_{n}|=1}\prod_{n=1}^{N}\frac{dz_{n}}{2\pi i z_{n}}\prod_{n\neq m}\bigg(1-\frac{z_{n}}{z_{m}}\bigg) \, \PE\bigg[\Isp^{\rm Maxwell}(q,u)\,\sum_{n,m=1}^{N}\frac{z_{n}}{z_{m}}\bigg],
\ee
where, \cf (\ref{2.16}), 
\be
\Isp^{\rm Maxwell}(q,u) = \frac{q(u+u^{-1})-2q^{2}}{1-q^{2}}\,,
\ee
and $z_{n}$ are gauge fugacities for the Maxwell multiplet in adjoint representation. In the $N=2$ case one of the two integrations trivializes and 
we can set from the start $z_{1}\to z$, $z_{2}\to 1$. This gives
\be
\la{B.3}
\I_{2}^{\rm SYM}(q,u) = \oint_{|z|=1}\frac{dz}{2\pi i} F(z; q,u), 
\ee
with 
\be
\la{B.4}
F(z; q,u) = \frac{1}{2z}\,\PE\bigg[2-\frac{(1-\frac{q}{u})(1-qu)}{(1-q^{2})}(z+2+z^{-1})\bigg]\, .
\ee
After some rearrangement  this can be written 
\be
\la{B.5}
F(z; q,u) = -\frac{(q^2;q^2)^4_{\infty} (\frac{1}{z};q^2)_{\infty}^2 
(z;q^2)_{\infty}^2}{2 (1-z)^2 (\frac{q}{u};q^2)_{\infty}^2 (q 
u;q^2)_{\infty}^2 (\frac{q}{u z};q^2)_{\infty} (\frac{q 
u}{z};q^2)_{\infty} (\frac{q z}{u};q^2)_{\infty} (q u 
z;q^2)_{\infty}}\, .
\ee
The  index of the D3 brane wrapped $k$ times over $\sX=0$ is given by (\ref{3.13}).
The problem is that if we start with a $q$-series for $\I_{k}^{\rm SYM}(q,u)$, 
the transformation 
\be
\la{B.6}
q\to q^{1/2}u^{-1/2}, \qquad u\to q^{-3/2}u^{-1/2}\,,
\ee
will convert it into a series with an infinite number of singular terms $1/q^{n}$ mixed with $1/u^{m}$ and these have to be resummed before re-expanding at small $q$ and $|u|<1$.
The strategy adopted in \cite{Arai:2020qaj} was based on applying (\ref{B.6}) to $F(z; q,u)$, expand in small $q$ and then deform the integration contour of $\{z_{n}\}$ in order to 
pick the relevant poles, no more in the circle $|z|=1$. In addition, when taking the $q$ expansion of $F(z; q^{1/2}u^{-1/2},q^{-3/2}u^{-1/2})$ some spurious poles collapse at $z=0$ 
and have to be subtracted by hand. Here, we adopt a direct approach and evaluate (\ref{B.3}) in closed form. The result will be expressed in terms of elliptic functions
and the continuation (\ref{B.6}) will be easy to perform exploiting their symmetries. After that, the small $q$ expansion can be safely done.

As a first step, we use the triple product  representation
\be
\la{B.7}
\vartheta_{4}(u,q^{1/2}) = (q,q)_{\infty}(x,q)_{\infty}(q/x,q)_{\infty}, \qquad x=q^{1/2}e^{2 iu}\,,
\ee
and the shift relation 
\be
\la{B.8}
\vartheta_{1}(u,q) = ie^{-iu}q^{1/4}\vartheta_{4}(u-\frac{\pi}{2}\tau, q)\,,
\ee
where our convention is $q=e^{i\pi \tau}$ as in (\ref{A.1}). If we parametrize 
%These relations allows to 
% bring $F(z; q,u)$ in the form 
%\be
%\la{B.9}
%F(z; q, u) = \frac{1}{2 \sqrt{q} z}\frac{(q^2,q^2)_{\infty}^4}{ (\frac{q}{u},q^2)_{\infty}^2 (q u,q^2)_{\infty}^2}
%\frac{\vartheta _1(\pi  x,q)^2 }{ \vartheta _4(\pi x-\pi  U,q) \vartheta _4(\pi x+\pi  U,q) }\, ,
%\ee
%where we parametrized
\be
\la{B.10}
z = e^{2\pi i x}, \qquad u = e^{2\pi i U}\,,
\ee
we get 
%The bunch of factors independent on $z$ can be also simplified using again (\ref{B.7}) and (\ref{A.3})
%\be
%\frac{1}{2 \sqrt{q}}\frac{(q^2,q^2)_{\infty}^4}{ (\frac{q}{u},q^2)_{\infty}^2 (q u,q^2)_{\infty}^2} = \frac{1}{2 \sqrt{q}}\frac{(q^2,q^2)_{\infty}^6}{\vartheta_{4}(\pi U,q)^{2}}= \frac{1}{2 q}\frac{\eta(\tau)^{6}}{\vartheta_{4}(\pi U,q)^{2}}\,.
%\ee
%In summary, we have shown that 
\be
\I_{2}^{\rm SYM}(q,u) = \frac{1}{2 q}\frac{\eta(\tau)^{6}}{\vartheta_{4}(\pi U,q)^{2}}\oint\frac{dz}{2\pi i z}\, 
\frac{\vartheta _1(\pi  x,q)^2 }{ \vartheta _4(\pi x-\pi  U,q) \vartheta _4(\pi x+\pi  U,q) }.
\ee
This integral has been evaluated in \cite{Hatsuda:2022xdv} (or using the methods of \cite{Pan:2021mrw}, who treated the very similar $SU(2)$ case)
and the results reads ($\vartheta_{a}'(u,q) = \partial_{u}\vartheta_{a}(u,q)$)
%
%and the explicit calculation 
%because the 
%function 
%\be
%G(x) = \frac{\vartheta _1(\pi  x,q)^2 }{ \vartheta _4(\pi x-\pi  U,q) \vartheta _4(\pi x+\pi  U,q) }\, , 
%\ee
%obeys
%\be
%G(x)=G(x+1)=G(x+\tau)\, , 
%\ee
%and defines an elliptic function to which we can apply their algorithms. The result reads  
\be
\I_{2}^{\rm SYM}(q,u) =  \frac{1}{2 q}\frac{\eta(\tau)^{3}}{\vartheta_{4}(\pi U,q)}\frac{\vartheta_{4}'(\pi  U,q)}{ \vartheta _1(2\pi  U,q)}\, .
\ee
As a check we can write in $u\to 1$ limit (or $U\to 0$)
\be
\I_{2}^{\rm SYM}(q,u) =  \frac{1}{\vartheta_{4}(0,q)}(1-4q^{3}+9q^{8}-16q^{15}+25q^{24}+\cdots) = \frac{1}{\vartheta_{4}(0,q)}\sum_{k=0}^{\infty}(-1)^{k}(k+1)^{2}q^{k(k+2)}\,,
\ee
in agreement with Eq.(21) of \cite{Bourdier:2015wda}.
To this expression, we can apply the analytic continuation (\ref{B.6}) and get, from (\ref{3.13}), 
\be
\I^{\rm D3}_{\sX, 2}(q,u) = \I_{2}^{\rm SYM}(q^{1/2}u^{-1/2}, q^{-3/2}u^{-1/2}) =
\frac{1}{2}(\tfrac{u}{q})^{3/8}(\tfrac{q}{u},\tfrac{q}{u})_{\infty}^3\frac{ \vartheta _4^{\prime }(2 \pi  U+\frac{3 \pi}{2}\tau',\frac{\sqrt{q}}{\sqrt{u}}) }
{ \vartheta _1(4 \pi  U+3 \pi \tau',\frac{\sqrt{q}}{\sqrt{u}}) \,
\vartheta _4(2 \pi  U+\frac{3 \pi}{2}\tau',\frac{\sqrt{q}}{\sqrt{u}})}\, , 
\ee
where $\tau'$ is such that $\sqrt{q}/\sqrt{u} = e^{i\pi \tau'}$. Now, we use the general formula
\be
\la{B.18}
\vartheta_{1}(u+m\pi \tau,q) = (-1)^{m}q^{-m}e^{-i(2mu+m(m-1)\pi \tau)}\vartheta_{1}(u,q)\, , 
\ee
to get 
\be
\vartheta_{1}(u+3\pi \tau,q) = -q^{-9}e^{-6iu}\vartheta_{1}(u,q)\, .
\ee
Also, from (\ref{B.8}) and (\ref{B.18}), one easily obtains
\ba
\vartheta_{4}(u+\frac{3\pi}{2}\tau,q) &= -i e^{-3iu}q^{-9/4}\vartheta_{1}(u,q)\, .
\ea
Differentiating with respect to $u$ gives also
\ba
\vartheta_{4}'(u+\frac{3\pi}{2} \tau,q) &=  -e^{-3iu}q^{-9/4}[3\vartheta_{1}(u,q)+i\vartheta_{1}'(u,q)]\, .
\ea
From these relations, we obtain the following closed formula for the 2-wrapped D3 brane index
\be
\I_{\sX, 2}^{\rm D3}(q,u) = \frac{i}{2} q^{33/8} u^{63/8}(\tfrac{q}{u},\tfrac{q}{u})_{\infty}^3\, \frac{ 3 \vartheta _1(2 \pi  
U,\frac{\sqrt{q}}{\sqrt{u}})+i \vartheta _1^{\prime }(2 \pi  U,\frac{\sqrt{q}}{\sqrt{u}}) }{
\vartheta _1(2 \pi  U,\frac{\sqrt{q}}{\sqrt{u}}) \vartheta _1(4 \pi  
U,\frac{\sqrt{q}}{\sqrt{u}})}
\ee
Expanding this in powers of $q$ gives the following terms
\ba
\I_{\sX, 2}^{\rm D3}(q,u) &= \frac{u^{10} (2-u^2) q^4}{(1-u^2)(1-u^{4})}+u^5 (2-u^4) q^5+(2+2 
u^6-u^{12}) q^6\lp
+\bigg(\frac{2}{u^5}+3 u^7-u^{15}\bigg) 
q^7+\bigg(3+\frac{2}{u^{10}}+3 u^8-u^{18}\bigg) 
q^8+\bigg(\frac{2}{u^{15}}+\frac{1}{u^3}+4 u^9-u^{21}\bigg) 
q^9\lp
+\bigg(3+\frac{2}{u^{20}}+\frac{4}{u^6}+4 u^{10}-u^{24}\bigg) q^{10}+\cdots\, , 
\ea
in full agreement with the third line in Eq.~(18) of \cite{Arai:2020qaj}.

\bibliography{BT-Biblio}
\bibliographystyle{JHEP-v2.9}
\end{document}